\documentclass[prd,twocolumn,showpacs,preprintnumbers,superscriptaddress,floatfix,nofootinbib]{revtex4-1}
\usepackage{graphicx,,booktabs}
\usepackage{epstopdf}
\graphicspath{{figures/}}
\usepackage{amssymb}
\usepackage{amsmath,txfonts}
\usepackage{graphicx}
\usepackage{dcolumn}
\usepackage{color}
\usepackage{bm}
\usepackage[colorlinks,
citecolor=blue,
anchorcolor=green,
menucolor=orange,
linkcolor=red,
filecolor=red,
runcolor=pink,
urlcolor=blue,
frenchlinks=red]{hyperref}
\usepackage{multirow}
\usepackage[capitalize]{cleveref}
\usepackage{slashed}
\usepackage{ulem}
\makeatletter

\newcommand{\Rmnum}[1]{\expandafter\@slowromancap\romannumeral #1@}

\makeatother

\begin{document}
	
	\title{\boldmath Production potential of hidden-strange molecular pentaquarks through the $\pi^-p\rightarrow K^{*}\Sigma$ process}

	\author{Xiao-Yun Wang}
	\email{xywang@lut.edu.cn}
	\affiliation{Department of physics, Lanzhou University of Technology,
		Lanzhou 730050, China}
	\affiliation{Lanzhou Center for Theoretical Physics, Key Laboratory of Theoretical Physics of Gansu Province, Lanzhou University, Lanzhou, Gansu 730000, China}
	
	\author{Yuan Gao}
	\affiliation{Department of physics, Lanzhou University of Technology,
		Lanzhou 730050, China}
	
	\author{Xiang Liu}
	\email{xiangliu@lzu.edu.cn}
	\affiliation{Lanzhou Center for Theoretical Physics, Key Laboratory of Theoretical Physics of Gansu Province, Lanzhou University, Lanzhou, Gansu 730000, China}
	\affiliation{School of Physical Science and Technology, Lanzhou University, Lanzhou 730000, China}
	\affiliation{Key Laboratory of Quantum Theory and Applications of MoE, Lanzhou University, Lanzhou 730000, China}
	\affiliation{MoE Frontiers Science Center for Rare Isotopes, Lanzhou University, Lanzhou 730000, China}
	\affiliation{Research Center for Hadron and CSR Physics, Lanzhou University and Institute of Modern Physics of CAS, Lanzhou 730000, China}
	
	\begin{abstract}
		In this study, applying the effective Lagrangian approach, we investigate the reaction $\pi^-p\rightarrow K^{*}\Sigma$ to explore the production of the hidden-strange molecular pentaquarks $P_{s\bar{s}}$. Specifically, we analyze two scenarios, where the $J^P$ quantum numbers of $P_{s\bar{s}}$ are $3/2^-$ and $1/2^-$. Our results show that the contribution from $s$-channel $P_{s\bar{s}}$ exchange dominates the total cross section, surpassing those from $t$-channel $K^{(*)}$ exchange and $u$-channel $\Sigma$ exchange. Additionally, we present predictions for the differential cross section of the $\pi^-p\rightarrow K^{*}\Sigma$ process. Finally, we extend the 2-to-2 scattering process to the 2-to-3 Dalitz process and provide theoretical predictions for this scenario. Our findings suggest that the cross section for the $\pi^-p \rightarrow K^{*}\Sigma\rightarrow K^+\pi^-\Sigma^0$ process can reach several tens of $\mu b$ near the threshold, making it highly favorable for experimental measurement. However, the current scarcity of experimental data for the $\pi^-p\rightarrow K^{*}\Sigma$ reaction at threshold energy limits our ability to determine the properties of the hidden-strange molecular pentaquarks $P_{s\bar{s}}$ and related physical quantities. Therefore, additional experimental measurements of the $\pi^-p \rightarrow K^{*}\Sigma$ reaction at threshold are strongly encouraged, and correlation studies could be pursued at facilities such as J-PARC, AMBER, and upcoming HIKE and HIAF meson beam experiments.
	\end{abstract}
	\maketitle

\section{INTRODUCTION}

Hadron spectroscopy offers an effective approach to deepen our understanding of the nonperturbative aspects of strong interactions. Over the past several decades, meson-nucleon scattering experiments have played a crucial role in the advancement of hadron spectroscopy \cite{wl:2024ext}. This is evident from the fact that over $80\%$ of the light hadrons cataloged by the Particle Data Group (PDG) \cite{ParticleDataGroup:2022pth} have been discovered through such scattering processes. While collider experiments are the primary means of exploring the nature of matter, scattering experiments deserve greater attention. In recent years, theorists have increasingly focused on scattering processes involving $K$ and $\pi$ meson beams, aiming to investigate the production of new hadronic states \cite{Liu:2019zoy,Chen:2016qju,Chen:2022asf,Zou:2016bxw,Lin:2022eau,Wang:2023lia,Wang:2022sib,Xiang:2020phx,Wang:2019zaw,Wang:2019uwk,Wang:2019dsi,Wang:2018mjz,Wang:2017qcw,Cheng:2016ddp,Wang:2015xwa,Liu:2021ojf,Gao:2012zh,Wu:2009tu,Xie:2014kja,Xie:2007qt}.

With the observation of several $P_c$ states by the LHCb Collaboration \cite{LHCb:2015yax,LHCb:2019kea}, hidden-charm molecular-type pentaquarks, predicted in Refs. \cite{Chen:2015loa,Chen:2015moa,Wang:2019krd,Shimizu:2016rrd,Roca:2015dva,Wang:2019dsi}, have been identified. This identification stems from the discovery of a characteristic mass spectrum in the $J/\psi p$ invariant mass distribution from the decay $\Lambda_b \to J/\psi p K$ \cite{LHCb:2015yax,LHCb:2019kea}. This represents an important progress in the study of hadron spectroscopy.

Following this, it is natural to further investigate the hidden-strange partners of these hidden-charm molecular-type pentaquarks while developing the ``Particle Zoo 2.0" version. Among the numerous observed light-flavor baryon states, the $N^*(2080)$ stands out as a particularly intriguing research target, closely linked to the investigation of hidden strange pentaquark states. The nucleon excited state $N^*(2080)$ has been proposed as a $K^*\Sigma$ molecular state, potentially serving as the strange partner of the $P_c^+(4457)$ hadronic molecular state~\cite{He:2017aps, Lin:2018kcc,Khemchandani:2011et,Doring:2008sv}. The potential existence of such hadronic molecules has also been suggested in studies utilizing the quark delocalization color screening model~\cite{Gao:2017hya}. In Ref.~\cite{Lin:2018kcc}, the effective Lagrangian method was employed to calculate the decay modes of the $N^*(2080)$ having $J^P=3/2^-$ as an $S$-wave $K^*\Sigma$ molecular state. The results showed that the measured decay properties of $N^*(2080)$ with $J^P=3/2^-$ were well reproduced, supporting the molecular interpretation of this state. We observe a change in the resonance designation relation to $N^*(2080)$.
According to the latest PDG~\cite{ParticleDataGroup:2022pth}, the previously identified two-star nucleon resonance $N^*(2080)$ has been split into a three-star $N(1875)$ and a two-star $N(2120)$, both with spin-parity $3/2^-$. Therefore, in our analysis, we no longer use $N^*(2080)$ to represent the hidden-strange molecular-type pentaquarks as considered in Refs. \cite{He:2017aps, Lin:2018kcc,Khemchandani:2011et,Doring:2008sv}. Instead, we focus on the production of hidden-strange molecular-type pentaquarks with the $K^*\Sigma$ 
component. This leads to two possible $J^P$ quantum numbers for S-wave $K^*\Sigma$ molecular systems: $J^P=1/2^-$ and $3/2^-$. In this study, we designate two S-wave $K^*\Sigma$ molecular-type pentaquarks with $J^P=1/2^-$ and $3/2^-$ as $P_{s\bar{s}}[1/2^-]$ and $P_{s\bar{s}}[3/2^-]$, respectively.

In this work, we investigate the production of two possible hidden-strange molecular-type pentaquarks
through the $\pi^{-} p\rightarrow K^{*}\Sigma$ reaction. We notice 
several previous experimental results, where the measured scattering cross section data for the $\pi^{-} p\rightarrow K^{*}\Sigma$ process were given~\cite{Crennell:1972km,CERN-CollegedeFrance-Madrid-Stockholm:1980mch,CERN-CollegedeFrance-Madrid-Stockholm:1980ysu,Dahl:1967pg,Miller:1965,Abramovich:1972rq}. It provides a good chance to combining with these experimental data to carry out the present research.
 
For quantitatively calculating the $\pi^{-} p\rightarrow K^{*}\Sigma$ scattering process, we adopt the effective Lagrangian approach, with the aim of investigating the production of hidden-strange molecular pentaquarks $P_{s\bar{s}}$. Additionally, we will analyze the $\pi^-p \rightarrow K^{*}\Sigma \rightarrow K^+\pi^-\Sigma^0$ Dalitz process and assess its experimental feasibility, thereby offering valuable theoretical insights for future experimental studies.

	This paper is organized as follows. After the introduction, the Lagrangians and amplitudes used in this work are presented in~\cref{zhangjie2}. The numerical results for the cross sections and the constituent counting rule are provided in \cref{zhangjie3}. The Dalitz process and its experimental feasibility are discussed in \cref{zhangjie5}, followed by a brief summary in \cref{zhangjie6}. 
	
	\section{THE PRODUCTION OF THE $P_{s\bar{s}}$ VIA THE $\pi^{-} p\rightarrow K^{*}\Sigma$ REACTION} \label{zhangjie2}

\begin{figure}[htbp]
		\centering
		\includegraphics[width=1.0\linewidth]{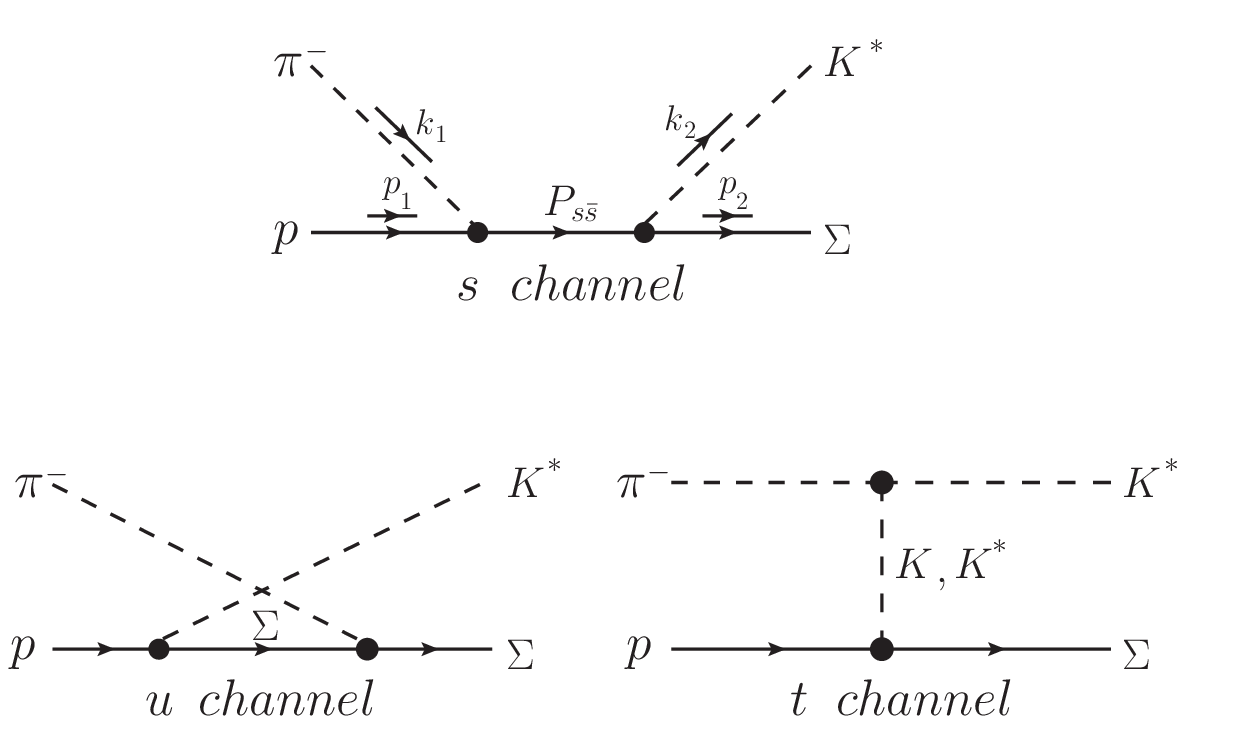}
		\caption{Feynman diagrams for the $\pi^-p\rightarrow K^{*}\Sigma$ reaction.}
		\label{fmt}
	\end{figure}

	The schematic tree-level Feynman diagrams for the $\pi^{-} p\rightarrow K^{*}\Sigma$ reaction are depicted in \cref{fmt}. These include $s$-channel $P_{s\bar{s}}$ ($\equiv P^{*}$) exchange, $u$-channel $\Sigma$ exchange, and $t$-channel $K$ and $K^*$ exchanges.
	
	To evaluate the pion-induced production of $K^*\Sigma$, the relevant Lagrangians are required, as used in previous studies~\cite{Xiang:2020phx,Kim:2015ita,He:2017aps,Xie:2014kja,Ben:2023uev,Wang:2019dsi,Cheng:2016ddp}. In general, similar to the case of the $P_c$ pentaquark state, the $J^P$ quantum numbers of the $P_{s\bar{s}}$ state can be either $\frac{3}{2}^-$ or $\frac{1}{2}^-$. Therefore, for the $s$-channel, both quantum number assignments for the $P_{s\bar{s}}$ should be considered in our calculations. These Lagrangians include 
	\begin{eqnarray}
		\mathcal{L}^{3/2^-}_{K^*\Sigma P^*}&=&g_{K^*\Sigma P^*}^{3/2^-}\bar{P^*}_{\mu}{\Sigma}K^{*\mu} +\mathrm{H.c.},\\
		\mathcal{L}^{3/2^-}_{\pi NP^*}&=&\frac{g_{\pi NP^*}^{3/2^-}}{m_\pi}\bar{N}\gamma_{5}\bm{\tau}\cdot\partial^{\mu}\bm{\pi}P^*_\mu + \mathrm{H.c.},\\
		\mathcal{L}^{1/2^-}_{K^*\Sigma P^*}&=&g_{K^*\Sigma P^*}^{1/2^-}\bar{P^*}\gamma_{5}\gamma^{\mu}K^*_\mu\Sigma + \mathrm{H.c.},\\
		\mathcal{L}^{1/2^-}_{\pi NP^*}&=&g_{\pi NP^*}^{1/2^-}\bar{N}\bm{\tau}\cdot\bm{\pi}P^* + \mathrm{H.c.}.
	\end{eqnarray}%
	The Lagrangians corresponding to the $u$-channel are written as
	\begin{eqnarray}
		\mathcal{L}_{\Sigma\Sigma \pi}&=&-\frac{g_{\Sigma\Sigma \pi}}{m_\pi}\bar\Sigma\gamma^{5}\gamma^{\mu}\partial_{\mu}\bm{\pi}\cdot\bm{T}\Sigma,\\
		\mathcal{L}_{K^*N{\Sigma} }
		&=&-g_{K^*N{\Sigma} }\bar{N}\Sigma(K\mkern-11mu/^*-\frac{k_{K^*N{\Sigma} }}{2m_N}\sigma_{\mu\nu}\partial^{\nu} K^{*\mu}) + \mathrm{H.c.},
	\end{eqnarray}%
	The effective Lagrangians for the $t$-channel involving $K$ and $K^*$ exchanges are given as follows:
	\begin{eqnarray}
		\mathcal{L}_{\pi KK^*}&=&g_{\pi KK^*}[\bar K(\partial^{\mu}\bm{\tau}\cdot\bm{\pi})-(\partial^{\mu}\bar K)\bm{\tau}\cdot\bm{\pi}]K^*_{\mu} + \mathrm{H.c.},\\
		\mathcal{L}_{\pi K^*K^*}&=&g_{\pi K^*K^*}\epsilon^{\mu \nu\alpha\beta}\partial_{\mu}\bar K^*_{\nu}\bm{\tau}\cdot\bm{\pi}\partial_{\alpha}K^*_\beta,\\
		\mathcal{L}_{KN\Sigma  } &=&g_{KN\Sigma  }\bar{N}\gamma_{5}K\bm{\tau}\cdot\bm{\Sigma} + \mathrm{H.c.},\\
		\mathcal{L}_{K^*N{\Sigma} }
		&=&-g_{K^*N{\Sigma} }\bar{N}\Sigma(K\mkern-11mu/^*-\frac{k_{K^*N{\Sigma} }}{2m_N}\sigma_{\mu\nu}\partial^{\nu} K^{*\mu}) + \mathrm{H.c.},
	\end{eqnarray}
	where $\bm{\tau}$ denotes the Pauli matrices, and the matrix $\bm{T}$ is defined as in Ref.~\cite{Matsuyama:2006rp}. The symbols $\bm{\pi}$, $K$, and $K^*$ represent the $\pi$, $K$, and $K^*$ meson fields, respectively. Similarly, $N$, $\Sigma$, and $P^*$ correspond to the nucleon, $\Sigma$ baryon, and $P_{s\bar{s}}$ fields, respectively. 
	
	Assuming the $P_{s\bar{s}}[3/2^-]$ is a pure $S$-wave molecular state of $K^*$ and $\Sigma$, the coupling constant $g_{K^*\Sigma P^*}^{3/2^-}$ can be estimated model-independently using the Weinberg compositeness criterion, yielding~\cite{Weinberg:1965zz,Baru:2003qq,Lin:2017mtz,Lin:2018kcc} 
	\begin{eqnarray}
		\label{Ampp}
		g_{K^*\Sigma P^*}^2&=&\frac{4\pi}{4M_{P^*}M_\Sigma}\frac{(M_{K^*}+M_\Sigma)^{5/2}}{(M_{K^*}M_\Sigma)^{1/2}}\sqrt{32\epsilon},
	\end{eqnarray}%
	\begin{eqnarray}
		\epsilon&\equiv&M_{K^*}+M_\Sigma-M_{P^*},
	\end{eqnarray}
	where $M_{P^*}$, $M_{K^*}$, and $M_\Sigma$ represent the masses of the $P_{s\bar{s}}[3/2^-]$, $K^*$, and $\Sigma$, respectively, while $\epsilon$ denotes the $K^*\Sigma$ binding energy. Assuming the physical state is a pure $S$-wave hadronic molecule, the relative uncertainty of the above approximation for the coupling constant is given by $\sqrt{2\mu\epsilon} \cdot r$, where $\mu = \frac{m_1m_2}{m_1+m_2}$ is the reduced mass of the bound particles, and $r$ is the range of forces, which can be estimated as the inverse of the mass of the exchanged particle. For the $K^*\Sigma$ system, $r$ can be estimated as $1/m_K$~\cite{Lin:2018kcc}.
	
	Following Ref.~\cite{Lin:2018kcc}, we adopt the mass of the $P_{s\bar{s}}[3/2^-]$ as $M_{P^*} = 2080 \text{ MeV}$. Substituting this into Eq.~(\ref{Ampp}), one obtains:
	\begin{eqnarray}
		g_{K^*\Sigma P^*}^{3/2^-}=1.72.
	\end{eqnarray}%
	Since the $P_{s\bar{s}}[1/2^-]$  can also be considered a $S$-wave molecular state of $K^*$ and $\Sigma$, we approximate the value of $g_{K^*\Sigma P^*}^{1/2^-}$ as 1.72. In the absence of experimental measurements for the partial width of the $P_{s\bar{s}}$ decays to $\pi N$, we treat the coupling constants $g_{\pi NP^*}^{3/2^-}$ and $g_{\pi NP^*}^{1/2^-}$ as free parameters to be determined through fitting.
	
	The coupling constant values for the $\Sigma\Sigma\pi$, $K^*N\Sigma$, and $KN\Sigma$ interactions have been provided in several theoretical works~\cite{Xiang:2020phx, He:2017aps, Cheng:2016ddp}. We use the following values: $g_{\Sigma\Sigma\pi} = 0.8$, $g_{K^*N\Sigma} = -2.46$, $\kappa_{K^*N\Sigma} = -0.47$, and $g_{KN\Sigma} = 2.69$.
	
	The coupling constant $g_{\pi KK^*}$ can be determined from the decay width $\Gamma_{K^* \to K \pi}$, as detailed in~\cite{Xiang:2020phx}
	\begin{eqnarray}
		{\Gamma}_{K^*{\rightarrow}K\pi}&=&\frac{g^{2}_{\pi KK^*}}{2\pi}\frac{|\vec{p}_\pi^{c.m.}|^3}{m^2_{K^*}}
	\end{eqnarray}
	with
	\begin{eqnarray}
		|\vec{p}_\pi^{c.m.}|&=&\frac{\sqrt{[m^2_{K^*}-(m_K+m_\pi)^2][m^2_{K^*}-(m_K-m_\pi)^2]}}{2m_{K^*}},	
	\end{eqnarray}
	where $\lambda$ is the K$\ddot{a}$llen function, defined as $\lambda(x,y,z)=(x-y-z)^2-4yz$. 
	The masses of the pion, $K$, and $K^*$ mesons are denoted by $m_{\pi}$, $m_{K}$, and $m_{K^*}$, respectively. Using the experimental data for the $K^*$ meson from PDG \cite{ParticleDataGroup:2022pth}, the coupling constant $g_{\pi KK^*}$ can be calculated. 
	With a mass of $m_{K^*} = 895.6 \text{ MeV}$ and a total decay width of $\Gamma_{K^*} = 47.3 \text{ MeV}$, along with the branching ratio $\mathrm{BR}(K^* \rightarrow K\pi) = 1.00$, we obtain $g_{\pi KK^*} = 3.10$. To determine the $\pi K^*K^*$ coupling constant $g_{\pi K^*K^*}$, we apply hidden local gauge symmetry \cite{Furui:1995bj} and flavor SU(3) symmetry, yielding a value of $g_{\pi K^*K^*} = 7.45$.

	For the $t$-channel and $u$-channel, the form factors are adopted in our calculation as follows,
	\begin{eqnarray}
		F_{t/u}(q) =\left(\frac{\Lambda{^2_{t/u}}-m^2}{\Lambda{^2_{t/u}}-q^2}\right)^2.
	\end{eqnarray}%
	
	To account for the finite size of the hadron, we incorporate a form factor in our calculation for the $s$-channel with an intermediate baryon, as follows:
	\begin{eqnarray}
		F_s(q) = \frac{\Lambda{^4_s}}{\Lambda{^4_s}+(q^2-m^2)^2},
	\end{eqnarray}%
	where $q$ and $m$ represent the four-momentum and mass of the exchanged particles, respectively. 
	
	%\subsection{Amplitudes}
	Based on the above Lagrangians, the scattering amplitude for the reaction $\pi^-p \rightarrow K^*\Sigma$ is given by:
	\begin{eqnarray}
		\mathcal{M}&=&\epsilon^{\mu}(k_2)\bar{u}(p_{2})(\mathcal{A}_{s, \mu}+\mathcal{A}_{u, \mu}+\mathcal{A}_{t, \mu}){u}(p_{1}),
	\end{eqnarray}
	where $\epsilon^{\mu}$ denotes the polarization vector of the outgoing $K^*$ meson, while $\bar{u}$ and $u$ represent the Dirac spinors for the outgoing $\Sigma$ baryon and the incoming proton, respectively.
	
	The reduced amplitudes $\mathcal{A}_{s, \mu}$, $\mathcal{A}_{u, \mu}$, and $\mathcal{A}_{t, \mu}$ for the $s$-, $u$-, and $t$-channel contributions read as
	\begin{eqnarray}
		\label{Amp}
		\mathcal{A}^{P^{*({3/2}^-)}}_{s,\mu} &=&\sqrt{2}g_{K^*\Sigma P^*}^{{3/2}^-}\frac{g_{\pi NP^*}^{{3/2}^-}}{m_\pi}F_s(q)\gamma_5k_{1}^\nu\notag \\ &&\times\frac{(q\mkern-8mu/_s+m_{P^*})}{s-m^2_{P^*}+im_{P^*}\Gamma_{P^*}}\Delta_{\mu\nu},\\
		\label{Amp4}
		\mathcal{A}^{P^{*({1/2}^-)}}_{s,\mu} &=&-i\sqrt{2}g_{K^*\Sigma P^*}^{{1/2}^-}g_{\pi NP^*}^{{1/2}^-}F_s(q)\gamma_5\gamma_\mu \nonumber\\ &&\times\frac{(q\mkern-8mu/_s+m_{P^*})}{s-m^2_{P^*}+im_{P^*}\Gamma_{P^*}},\\
		\label{Amp1}
		\mathcal{A}^{\Sigma}_{u,\mu} &=&-i\sqrt{2}\frac{g_{\Sigma\Sigma\pi}}{m_\pi}g_{K^*N\Sigma }F_u(q)\gamma^{5}\gamma^{\nu}k_{1\nu}\frac{(q\mkern-8mu/_\Sigma+m_{\Sigma})}{{u-m_{\Sigma}^2}}\notag \\
		&&\times(\gamma_{\mu}-\frac{k_{K^*{\Sigma}N }}{2m_N}\gamma_{\mu}q\mkern-8mu/_{K^*}), \\ 
		\label{Amp2}
		\mathcal{A}^{K}_{t,\mu} &=&i\sqrt{2}g_{K^*K\pi}g_{K\Sigma N}F_t(q)\frac{1}{{t-m_{K}^2}}\gamma_{5}\notag \\
		&&\times[k_{1\mu}+(k_1-k_2)_\mu], \\ 
		\label{Amp3} 
		\mathcal{A}^{K^*}_{t,\mu} &=&i\sqrt{2}g_{\pi K^*K^*}g_{K^*N\Sigma}F_t(q)(\gamma_{\xi}-\frac{k_{K^*{\Sigma}N }}{2m_N}\gamma_{\xi}q\mkern-8mu/_{K^*})
		\notag \\
		&&\times \epsilon_{\mu\nu\alpha\beta}\frac{\mathcal{P}^{\nu\xi}}{{t-m_{K^*}^2}}k_{2}^\alpha(k_1-k_2)^{\beta}   
	\end{eqnarray}%
	with
	\begin{eqnarray}
		\mathcal{P}^{\nu\xi}&=&i(g^{\nu\xi}+q_{K^*}^\nu q_{K^*}^\xi/{m_{K^*}^2}),
	\end{eqnarray}%
	\begin{eqnarray}
		\Delta_{\mu\nu}&=&-g_ {\mu\nu}+\frac{1}{3}\gamma_\mu\gamma_\nu \notag\\
		&& +\frac{1}{3m_{N^*}} (\gamma_\mu q_\nu-\gamma_\nu q_\mu) +\frac{2}{3m_{N^*}^2}q_\mu q_\nu,
	\end{eqnarray}%
	where $s=(k_1+p_1)^2$, $u =(p_2- k_1)^2$, and $t =(k_1- k_2)^2$ are the Mandelstam
	variables.
	
	%\subsection{Reggeized treatment}
	The Regge trajectory model has been successful in analyzing hadron production at intermediate and high energies \cite{Wang:2015hfm, Ozaki:2009wp, Wang:2015zcp, Wang:2017qcw, Storrow:1983ct}. In this model, Reggeization is achieved by replacing the $t$-channel propagator in the Feynman amplitudes (Eqs. (\ref{Amp2}) and (\ref{Amp3})) with the Regge propagator
	\begin{eqnarray}
		\frac{1}{t-m_K^2}\rightarrow(\frac{s}{s_{scale}})^{\alpha_K(t)}\frac{\pi\alpha^{\prime }_{K}}{\Gamma[1+{\alpha_K(t)}]\sin[\pi{\alpha_K(t)}]}, \\
		\frac{1}{t-m_{K^*}^2}\rightarrow(\frac{s}{s_{scale}})^{\alpha_{K^*}(t)-1}\frac{\pi\alpha^{\prime }_{K^*}}{\Gamma[{\alpha_{K^*}(t)}]\sin[\pi{\alpha_{K^*}(t)}]}.
	\end{eqnarray}%
	The scale factor $s_{\text{scale}}$ is fixed at 1 GeV. The Regge trajectories for $\alpha_K(t)$ and $\alpha_{K^*}(t)$ are given as \cite{Ozaki:2009wp}: 
	\begin{eqnarray}
		\alpha_K(t)=0.70(t - m_K^2), \alpha_{K^*}(t)=1 + 0.85(t - m_{K^*}^2).
	\end{eqnarray}%
	Notably, no additional free parameters are introduced with the incorporation of the Regge model.
	%-----------------------------------------------------------------------------
	\section{Numerical results}\label{zhangjie3}
	After completing the preparation, the differential cross section for the reaction $\pi^{-} p\rightarrow K^{*}\Sigma$ can be calculated and used for correlation analysis with experimental data. In the center of mass (c.m.) frame, the differential cross section is given by:
	\begin{eqnarray}
		\label{Aamp}
		\frac{d\sigma}{d \cos\theta} = \frac{1}{32\pi s}\frac{|\vec{k}_2^{c.m.}|}{|\vec{k}_1^{c.m.}|}\left(\frac{1}{2}\sum\limits_\lambda|\mathcal{M}|^2 \right),
	\end{eqnarray}%
	where $\theta$ is the angle between the
	outgoing $K^{*}$ meson and the direction of the $\pi$ beam
	in the c.m. frame. $\vec{k}_1^{c.m.}$ and $\vec{k}_2^{c.m.}$ denote the three-momenta of the initial $\pi$ beam and the final $K^{*}$ meson, respectively.
	
	\subsection{Production of the $P_{s\bar{s}}$  with $J^P$ = $\frac{3}{2}^-$}
	%\subsubsection{Fitting procedure}
	
	Using the MINUIT code from CERNLIB, we will fit the experimental data \cite{CERN-CollegedeFrance-Madrid-Stockholm:1980ysu, CERN-CollegedeFrance-Madrid-Stockholm:1980mch, Crennell:1972km, Abramovich:1972rq, Dahl:1967pg, Miller:1965} for the $\pi^{-} p\rightarrow K^{*}\Sigma$ reaction. Both total and differential cross section datas are used in a $\chi^2$ fitting algorithm to determine the values of the free parameters. The fitting parameters are listed in Table \ref{width}.
	The fit excluding the contribution of the $P_{s\bar{s}}[3/2^-]$ yields $\chi^2/d.o.f.$ = 6.19, while including the $P_{s\bar{s}}[3/2^-]$ improves the fit to $\chi^2/d.o.f.$ = 1.52. This indicates a significant contribution from the $s$-channel $P_{s\bar{s}}[3/2^-]$.
	
\begin{figure}[htbp]
		\centering
		\includegraphics[scale=0.46]{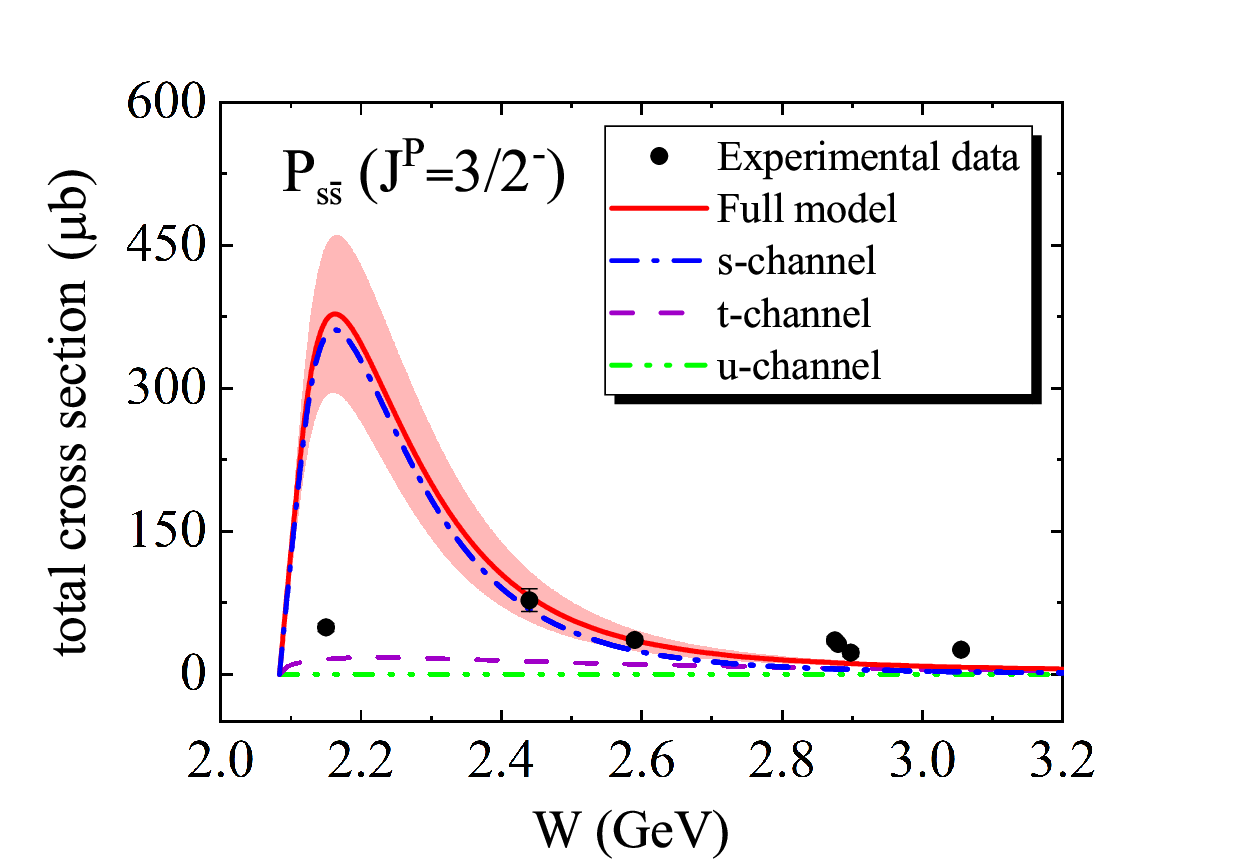}
		\caption{The total cross section for the reaction $\pi^-p\rightarrow K^{*}\Sigma$. The band stands for the error bar of the five fitting parameters in Table \ref{width}. The solid (red), dashed-dotted (blue), dashed (purple), dash–double dotted (green) lines are for the full model, the $s$-channel, the $t$-channel and the
			$u$-channel, respectively. Here, the spin-parity quantum number of the $P_{s\bar{s}}$ is $\frac{3}{2}^-$.}
		\label{tcs}
	\end{figure}

	\begin{table}[h]
		\renewcommand\arraystretch{1.5} 
		\caption{ Fitted values of the free parameters with all experimental data in Ref. \cite{CERN-CollegedeFrance-Madrid-Stockholm:1980ysu,CERN-CollegedeFrance-Madrid-Stockholm:1980mch,Crennell:1972km,Abramovich:1972rq,Dahl:1967pg,Miller:1965} for the case of the $P_{s\bar{s}}$ with $J^P=\frac{3}{2}^-$.  }
		\label{width}{\footnotesize \centering
			%\begin{tabular}{\linewidth}{ccc}
			\setlength{\tabcolsep}{4mm}{
				\begin{tabular}{ccc}
					\hline\hline
					& with $P_{s\bar{s}}[3/2^-]$ & without $P_{s\bar{s}}[3/2^-]$ \\
					\hline
					$\Lambda_K$  & 1.53$\pm$0.04&1.90$\pm$0.01\\
					\hline
					$\Lambda_{K^*}$ & 1.63$\pm$0.05&1.85$\pm$0.07\\
					\hline
					$\Lambda_u$ & 1.19$\pm$0.05&1.13$\pm$0.01\\
					\hline
					$\Lambda_s$ & 1.18$\pm$0.06&\\
					\hline
					$g_{\pi N{P^*}}^{{3/2}^-}$ & 0.37$\pm$0.04&\\
					\hline
					$\chi^2/d.o.f.$ & 1.52&6.19\\
					\hline\hline
			\end{tabular}}
		}
	\end{table}
	%\subsubsection{Cross section for the $\pi^{-} p\rightarrow K^{*}\Sigma$ reaction}
	The total cross
	section for the reaction $\pi^{-} p\rightarrow K^{*}\Sigma$ with the $P_{s\bar{s}}$ having $J^P$ = $\frac{3}{2}^-$ is illustrated in \cref{tcs}. The results, which include the contributions from the $t$-channel, $u$-channel, and $s$-channel, show good agreement with the experimental data. Additionally, \cref{tcs} demonstrates that while the contributions from the 
	$t$- and 
	$u$-channels are relatively small, the 
	$s$-channel plays a crucial role in the reaction.
	
	\begin{figure}[htbp]
		\centering
		\includegraphics[scale=0.46]{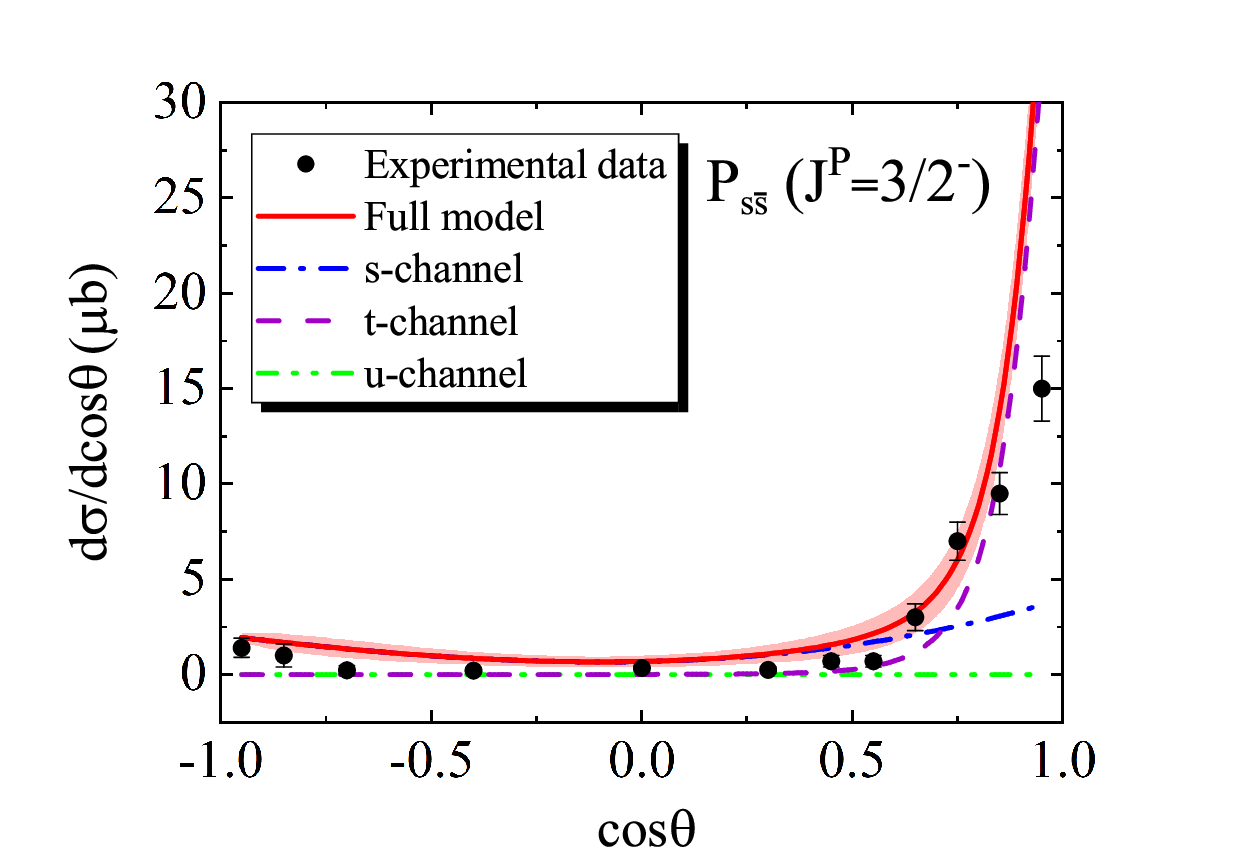}
		\caption{The differential cross-section $d\sigma/d cos\theta$ for the $\pi^-p\rightarrow K^{*}\Sigma$ reaction as a function of $\cos \theta$. The experimental datas are from Ref. \cite{Crennell:1972km}. Here, the notation is the same as that in \cref{tcs}.}
		\label{cos}
	\end{figure}
\begin{figure}[htbp]
		\centering
		\includegraphics[scale=0.46]{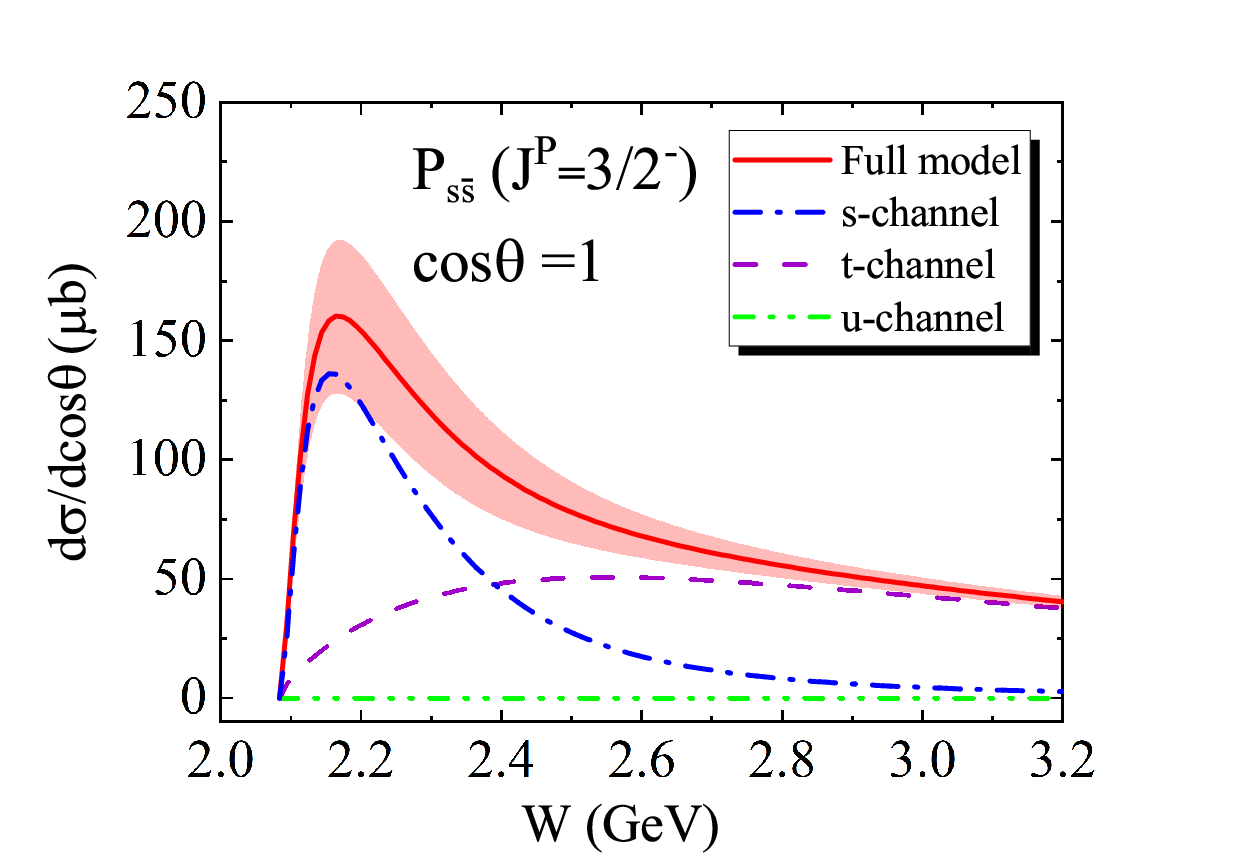}
		\caption{The differential cross-section $d\sigma/d \cos \theta$ for the $\pi^-p\rightarrow K^{*}\Sigma$ reaction varies with different c.m. energies when $\cos \theta$ = 1. Here, the notation is the same as that in \cref{tcs}.}
		\label{cos132}
	\end{figure}

	In \cref{cos}, we present the differential cross section for the reaction $\pi^- p \rightarrow K^* \Sigma$ as a function of $\cos \theta$, specifically for the $P_{s\bar{s}}$ with $J^P = \frac{3}{2}^-$ at a center-of-mass energy $W = 3.056$ GeV. The analysis reveals that the $t$-channel contribution dominates at forward angles, while the $u$-channel contribution is small and negligible. In \cref{cos132}, we show the differential cross section for the $\pi^- p \rightarrow K^* \Sigma$ reaction as a function of center-of-mass energy, with $\cos \theta = 1$, for the same $P_{s\bar{s}}$ with $J^P = \frac{3}{2}^-$. Since these measurements are taken at forward angles, the distinct shapes of the curves resulting from the contributions of the three channels allow us to effectively isolate the contributions from the $t$-channel and $u$-channel, while providing a clear indication of the $s$-channel $P_{s\bar{s}}[3/2^-]$. This approach is ideal for studying the contribution of the $P_{s\bar{s}}[3/2^-]$ through differential cross sections at forward angles.
	
\begin{figure}[htbp]
		\centering
		\includegraphics[scale=0.43]{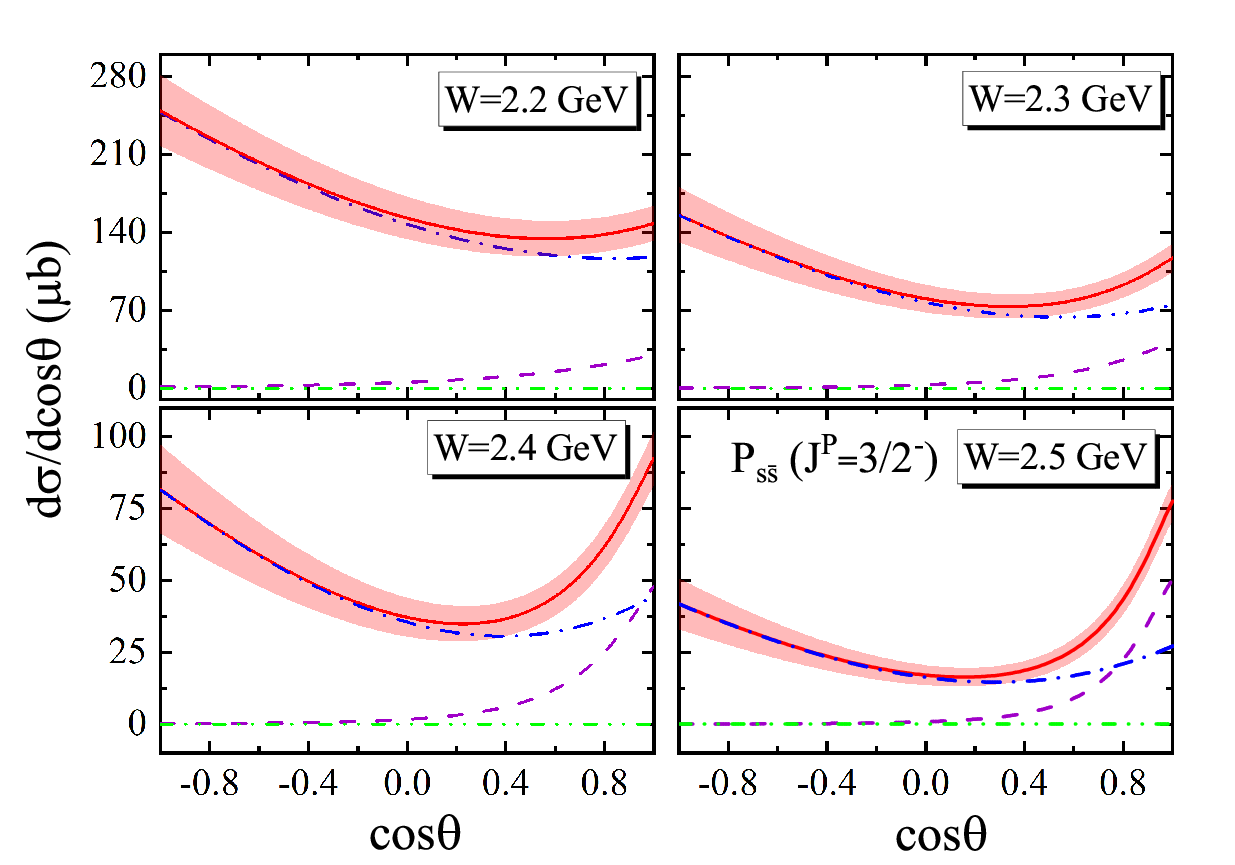}
		\caption{The differential cross section $d\sigma/d \cos \theta$ of the $\pi^-p\rightarrow K^{*}\Sigma$ process as a function of $\cos \theta$ at different c.m. energies. Here, the notation is the same as that in \cref{tcs}.}
		\label{cos32}
	\end{figure}

\begin{figure}[htbp]
		\centering
		\includegraphics[scale=0.46]{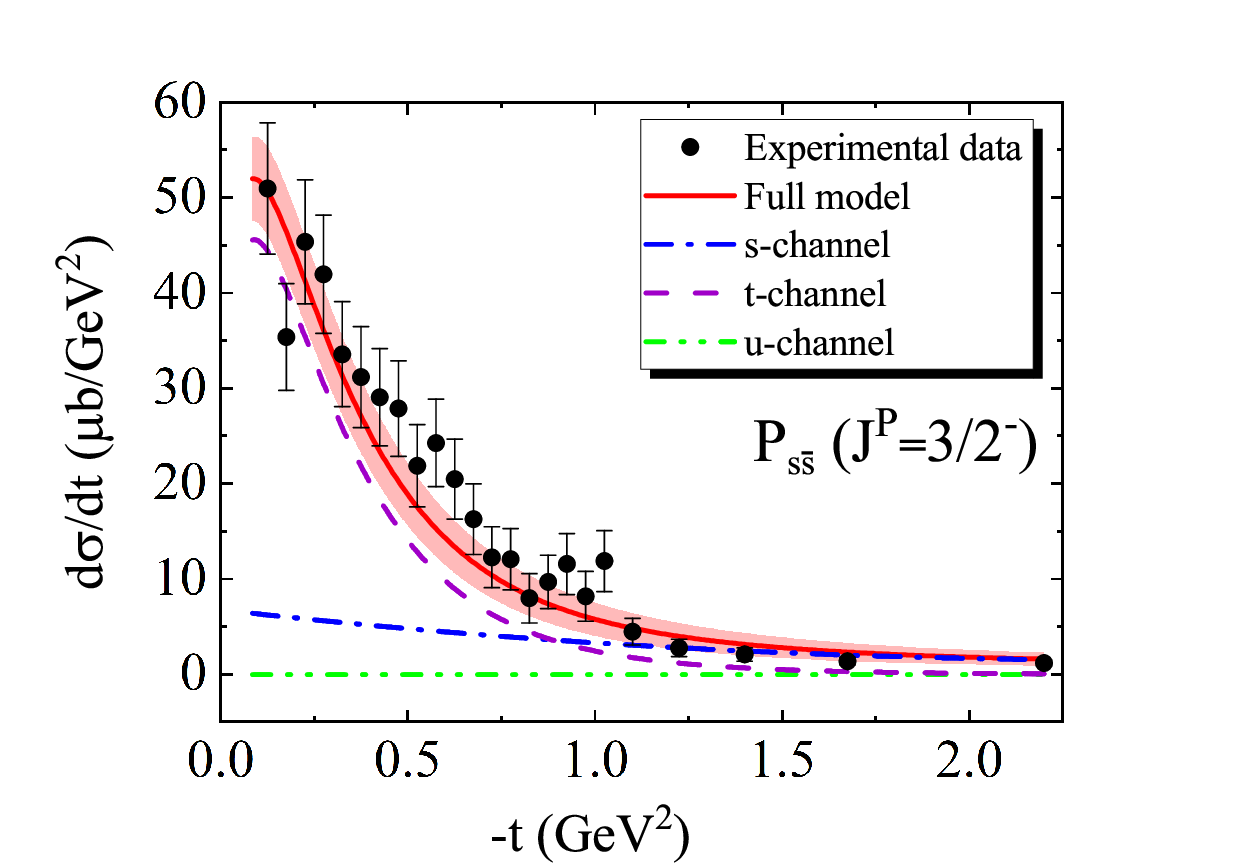}
		\caption{The $t$ distribution for the reaction  $\pi^-p\rightarrow K^{*}\Sigma$. The experimental datas are from Ref. \cite{CERN-CollegedeFrance-Madrid-Stockholm:1980mch}. Here, the notation is the same as that in \cref{tcs}.
			%The full (red), dashed (green), dashed-dotted (magenta), and dash-double dotted (blue) lines are the full model, $t$-, $s$-, acn
		}
		\label{t}
	\end{figure}

\begin{figure}[htbp]
		\centering
		\includegraphics[scale=0.43]{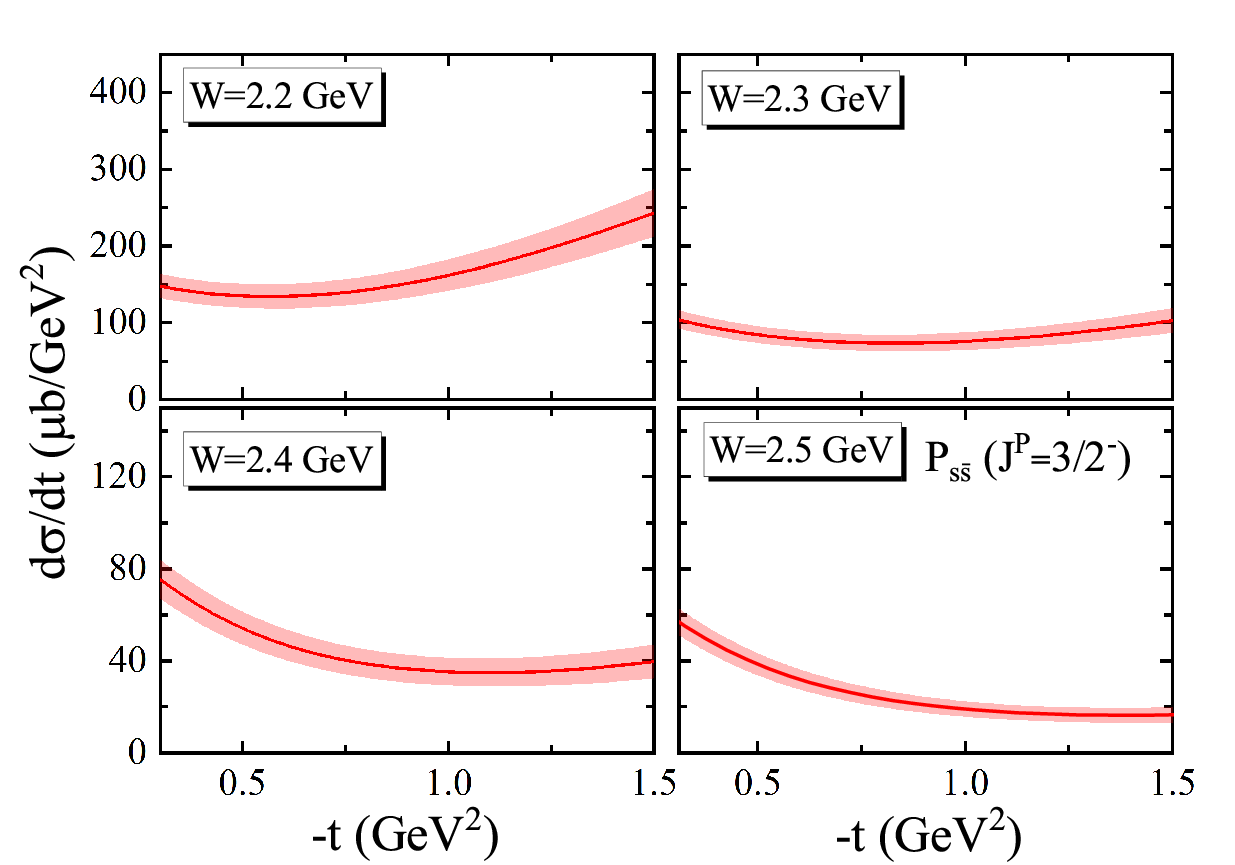}
		\caption{The $t$-distribution for the $\pi^-p\rightarrow K^{*}\Sigma$ reaction at different c.m. energies $W$ = 2.2 GeV, 2.3 GeV, 2.4 GeV and 2.5 GeV. Here, the notation is the same as in \cref{tcs}. }
		\label{t32}
	\end{figure}

	In \cref{cos32}, we calculate the differential cross section of the $\pi^- p \rightarrow K^* \Sigma$ reaction at various center-of-mass energies. The results indicate that as the energy approaches the threshold of $K^*\Sigma$, the contribution of the $s$-channel becomes more prominent. Conversely, as energy increases, the contribution of the $t$-channel at forward angles becomes more significant, while the $u$-channel contribution remains consistently small. These findings suggest that measuring the differential cross section of the reaction $\pi^- p \rightarrow K^* \Sigma$ near the $K^*\Sigma$ threshold will provide valuable information about the $P_{s\bar{s}}[3/2^-]$ resonance.   
	%\subsubsection{t distribution for the $\pi^{-} p\rightarrow K^{*}\Sigma$ reaction}
	The $t$-distribution for the reaction $\pi^{-} p \rightarrow K^{*} \Sigma$ with the $P_{s\bar{s}}$ and $J^P = \frac{3}{2}^-$ at $W = 2.88$ GeV is shown in \cref{t}. The results, incorporating full contributions from the $s$-channel, $t$-channel, and $u$-channel, are in good agreement with the experimental data. \cref{t32} presents the $t$-distribution for the $\pi^- p \rightarrow K^* \Sigma$ reaction at various center-of-mass energies. These results provide insight into the production mechanism of the $\pi^{-} p \rightarrow K^{*} \Sigma$ reaction, aiding in the differentiation of its underlying processes.
	
	\subsection{Production of the $P_{s\bar{s}}$ with $J^P$ = $\frac{1}{2}^-$}
	
\begin{figure}[htbp]
		\centering
		\includegraphics[scale=0.46]{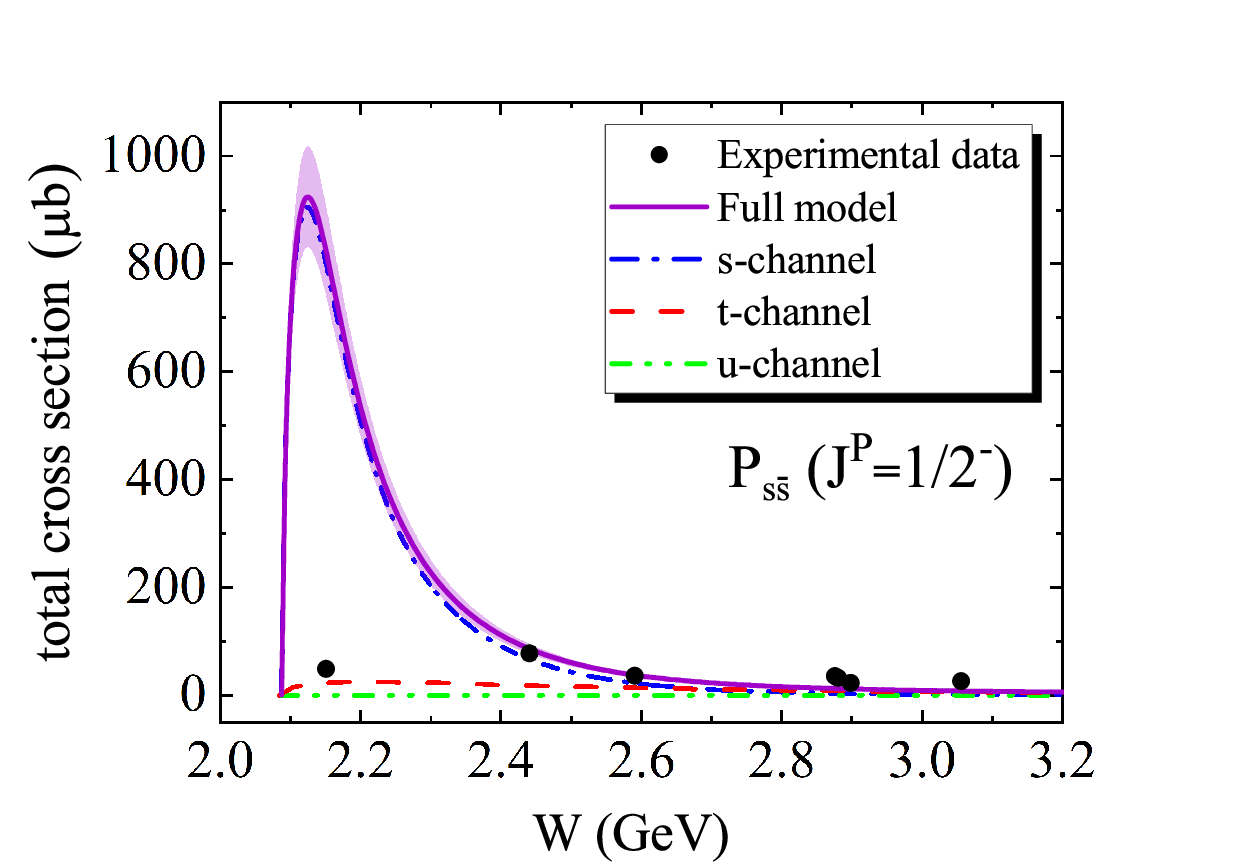}
		\caption{The total cross section for the reaction $\pi^-p\rightarrow K^{*}\Sigma$. The band stands for the error bar of the five fitting parameters in Table \ref{fit}. The solid (purple), dashed-dotted (blue), dashed (red), dash–double dotted (green) lines are for the full model, the $s$-channel, the $t$-channel and the
			$u$-channel, respectively. Here, the spin-parity quantum number of the $P_{s\bar{s}}$ is $\frac{1}{2}^-$.}
		\label{tcs2}
	\end{figure}

\begin{figure}[htbp]
		\centering
		\includegraphics[scale=0.46]{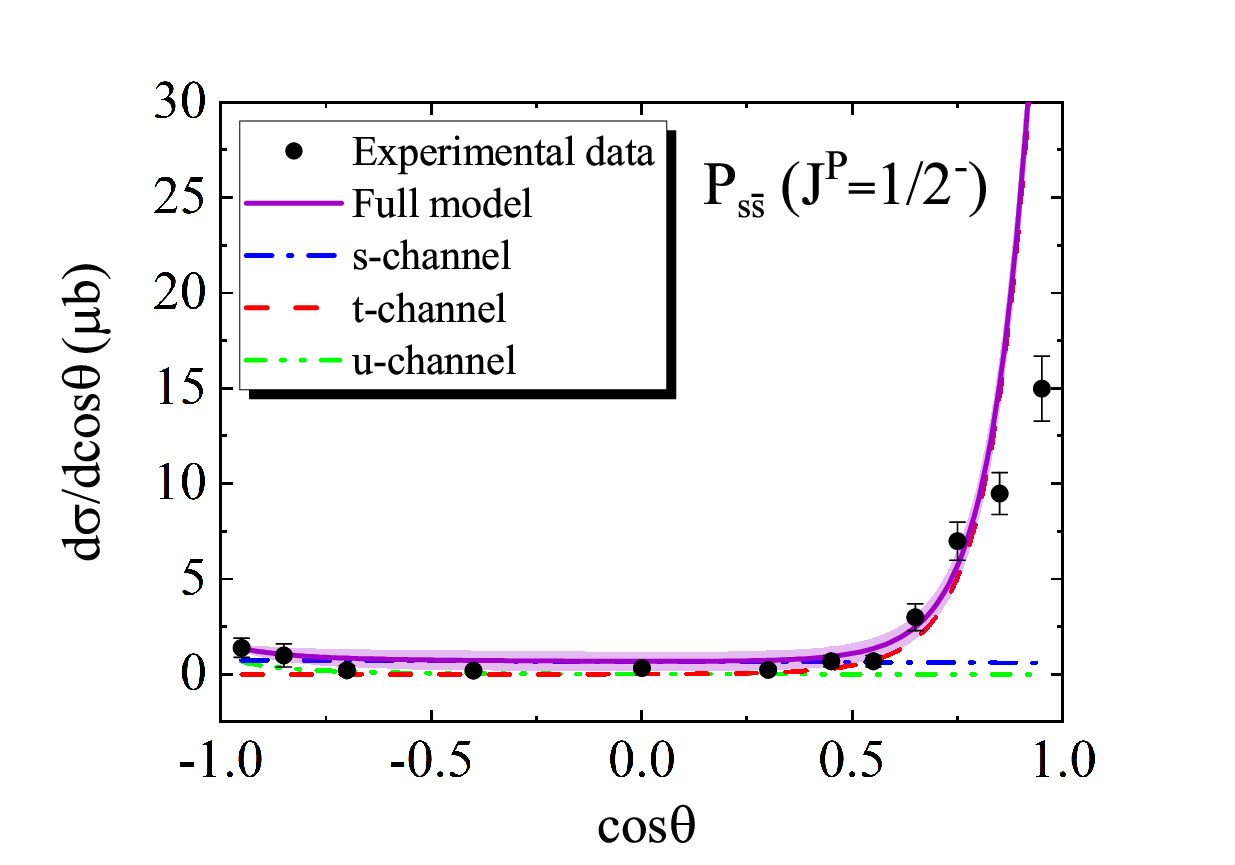}
		\caption{The differential cross-section $d\sigma/d \cos \theta$ for the $\pi^-p\rightarrow K^{*}\Sigma$ reaction as a function of $\cos \theta$. The experimental datas are from Ref. \cite{Crennell:1972km}. Here, the notation is the same as that in \cref{tcs2}.}
		\label{cos2}
	\end{figure}

\begin{figure}[htbp]
		\centering
		\includegraphics[scale=0.46]{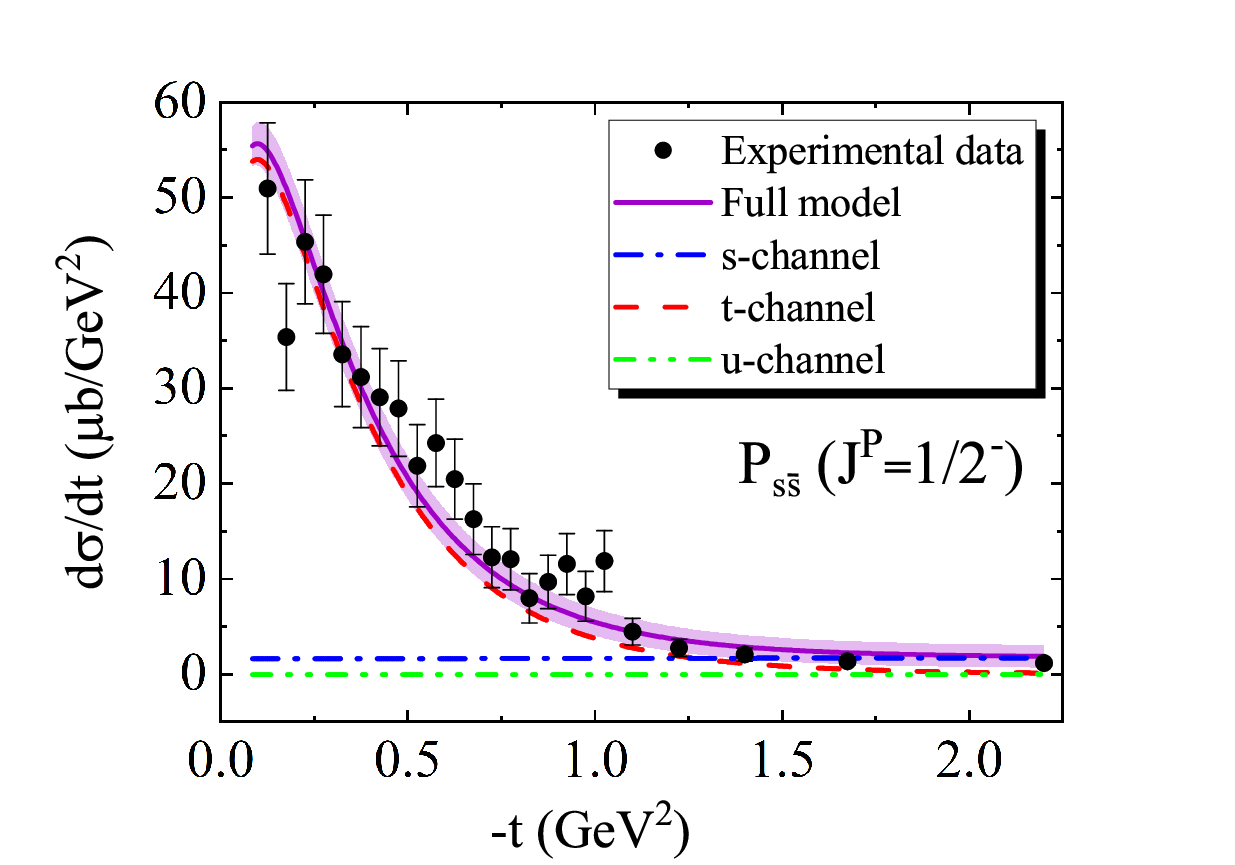}
		\caption{The $t$ distribution for the reaction  $\pi^-p\rightarrow K^{*}\Sigma$. The experimental datas are from Ref. \cite{CERN-CollegedeFrance-Madrid-Stockholm:1980mch}. Here, the notation is the same as that in \cref{tcs2}.}
		\label{t2}
	\end{figure}

	Similar to the analysis of $P_{s\bar{s}}$ with $J^P$ = $\frac{3}{2}^-$, we also calculate the cross section for the reaction $\pi^-p\rightarrow K^{*}\Sigma$ with $P_{s\bar{s}}$ having $J^P$ = $\frac{1}{2}^-$ as detailed in \cref{tcs2,cos2,t2}. 
	We employ the same fitting scheme as in Table \ref{width}, treating the cutoff value as a free parameter. The fitting parameters are shown in Table \ref{fit}. The fit excluding the $P_{s\bar{s}}[1/2^-]$ contribution yields a $\chi^2/d.o.f.$ = 6.19, whereas including this contribution results in $\chi^2/d.o.f.$ = 2.20. For the $P_{s\bar{s}}$ with $J^P$ = $\frac{3}{2}^-$, the $\chi^2/d.o.f.$ = 1.52, indicating a better fit to the experimental data for the $P_{s\bar{s}}[3/2^-]$. 
	
\begin{figure}[htbp]
		\centering
		\includegraphics[scale=0.46]{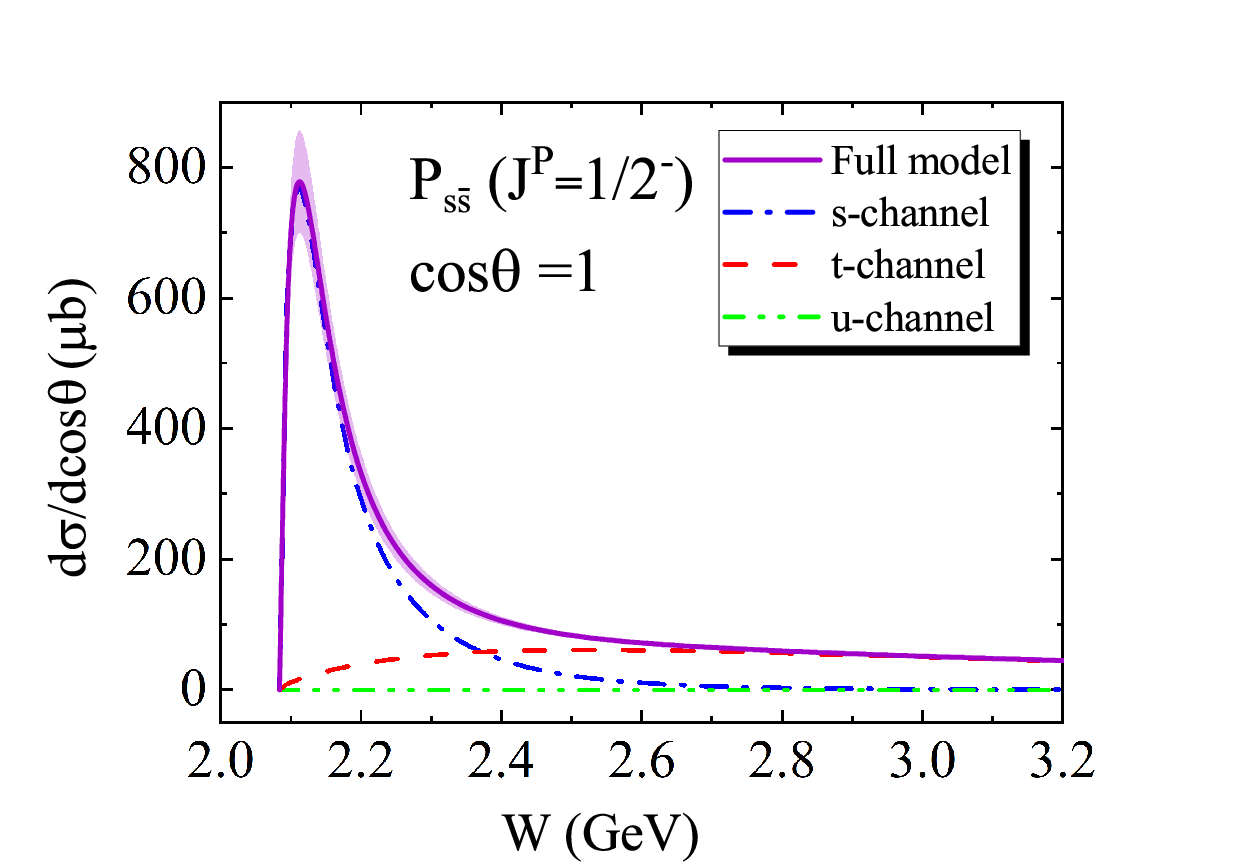}
		\caption{The differential cross-section $d\sigma/d \cos \theta$ for the $\pi^-p\rightarrow K^{*}\Sigma$ reaction varies with different c.m. energies when $\cos \theta$ = 1. Here, the notation is the same as that in \cref{tcs2}.}
		\label{cos112}
	\end{figure}

\begin{figure}[htbp]
		\centering
		\includegraphics[scale=0.43]{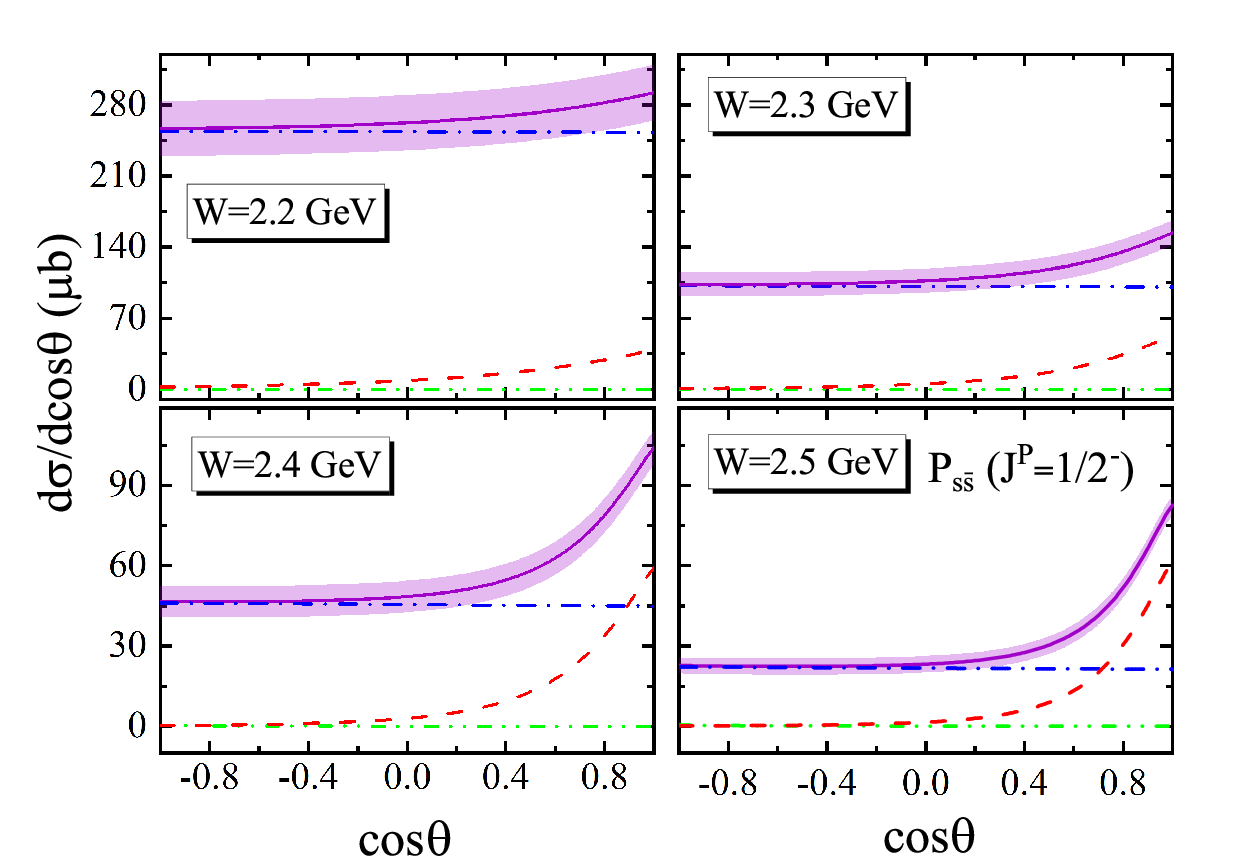}
		\caption{The differential cross section $d\sigma/d \cos \theta$ of the $\pi^-p\rightarrow K^{*}\Sigma$ process as a function of $\cos \theta$ at different c.m. energies. Here, the notation is the same as that in \cref{tcs2}.}
		\label{cos12}
	\end{figure}

\begin{figure}[htbp]
		\centering
		\includegraphics[scale=0.43]{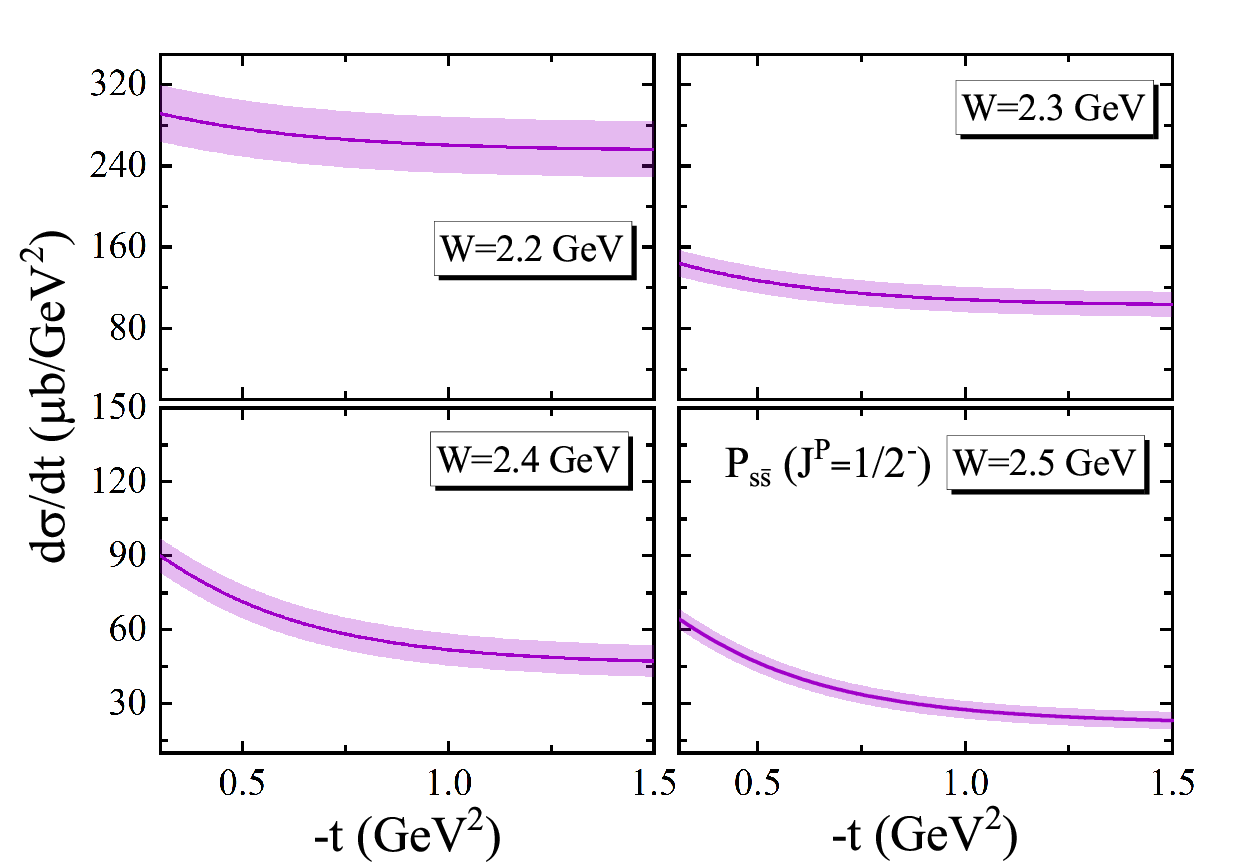}
		\caption{The $t$-distribution for the $\pi^-p\rightarrow K^{*}\Sigma$ reaction at different c.m. energies $W$ = 2.2 GeV, 2.3 GeV, 2.4 GeV and 2.5 GeV. Here, the notation is the same as in \cref{tcs2}. }
		\label{t12}
	\end{figure}

	\cref{cos2} and \cref{t2} present the differential cross section $d\sigma/d \cos \theta$ and the $t$-distribution for $\pi^-p\rightarrow K^{*}\Sigma$, respectively. These results closely match those in \cref{cos} and \cref{t}. \cref{tcs2} shows the total cross section for $\pi^-p\rightarrow K^{*}\Sigma$, with a peak value approximately 500 $\mu b$ higher than in \cref{tcs}. In \cref{cos112}, the differential cross section for the $\pi^-p\rightarrow K^{*}\Sigma$ reaction as a function of c.m. energies at $\cos \theta$ = 1 is presented. This provides an ideal means to assess the contribution of the $P_{s\bar{s}}[1/2^-]$ through forward-angle differential cross sections. \cref{cos12} shows the differential cross section at various c.m. energies. As the energy approaches the $K^*\Sigma$ threshold, the $s$-channel contribution becomes more pronounced. With increasing energy, the 
	$t$-channel contribution at forward angles grows more significant, while the $u$-channel contribution remains small. This behavior is similar to that observed for the $P_{s\bar{s}}[3/2^-]$. 
	\cref{t12} displays the $t$-distribution for the $\pi^-p\rightarrow K^{*}\Sigma$ reaction at different c.m. energies. These results may aid us in distinguishing the production mechanisms for the $\pi^{-} p\rightarrow K^{*}\Sigma$ reaction with the $P_{s\bar{s}}$ having $J^P$ = $\frac{1}{2}^-$. 
	\begin{table}[h]
		\renewcommand\arraystretch{1.5} 
		\caption{ Fitted values of the free parameters with all experimental data in Refs.~\cite{CERN-CollegedeFrance-Madrid-Stockholm:1980ysu,CERN-CollegedeFrance-Madrid-Stockholm:1980mch,Crennell:1972km,Abramovich:1972rq,Dahl:1967pg,Miller:1965} for the case of the $P_{s\bar{s}}$ with $J^P=\frac{1}{2}^-$.}
		\label{fit}{\footnotesize \centering
			%\begin{tabular}{\linewidth}{ccc}
			\setlength{\tabcolsep}{4mm}{
				\begin{tabular}{ccc}
					\hline\hline
					& with $P_{s\bar{s}}[1/2^-]$ & without $P_{s\bar{s}}[1/2^-]$ \\
					\hline
					$\Lambda_K$  &1.81 $\pm$0.02&1.90$\pm$0.01\\
					\hline
					$\Lambda_{K^*}$ &1.71$\pm$0.02&1.85$\pm$0.07\\
					\hline
					$\Lambda_u$ &1.14$\pm$0.01&1.13$\pm$0.01\\
					\hline
					$\Lambda_s$ &1.61$\pm$0.02&\\
					\hline
					$g_{\pi N{P^*}}^{1/2^-}$ &0.20$\pm$0.01&\\
					\hline
					$\chi^2/d.o.f.$ &2.20 &6.19\\
					\hline\hline
			\end{tabular}}
		}
	\end{table}
	

 \subsection{A further test by the constituent counting rule}\label{zhangjie4}
	In the following, we adopt the approach of the constituent counting rule to further test our theoretical calculation around the $\pi^- p \rightarrow K^*\Sigma$ reaction. Generally, for a large-angle exclusive scattering process $a+b \rightarrow c+d$, the reaction cross section can be expressed with the constituent counting rule (see Refs. \cite{Lepage:1980fj,Mueller:1981sg,Lepage:1979zb,Efremov:1979qk,Brodsky:1973kr,Brodsky:1974vy,Matveev:1973ra,Efremov:1978rn,Duncan:1979hi,Duncan:1980qd,Lepage:1980fj,Farrar:1979aw,Kawamura:2013iia, Anderson:1976ph, JeffersonLabHallA:2002vjd,JeffersonLabHallA:2004gxp,Baller:1988tj, White:1994tj} for more details):
	\begin{eqnarray}
		\label{13}
		\frac{d\sigma_{ab\to cd}}{dt}=\frac{1}{s^{n-2}}f_{ab\to cd}(t/s),
	\end{eqnarray}
	where $n\equiv n_a + n_b + n_c + n_d$, while $f(t/s)$ accounts for the scattering angle and is multiplied by the normalization factors. 
	\begin{table}[h]
		\renewcommand\arraystretch{1.5} 
		\caption{The cross sections of $\pi^-p\rightarrow K^{*}\Sigma$ at $\theta_{\mathrm{c.m.}}=90^\circ$ for the case of the $P_{s\bar{s}}$ with $J^P$ = $\frac{3}{2}^-$.}
		\label{32}{\footnotesize \centering
			%\begin{tabular}{\linewidth}{ccc}
			\setlength{\tabcolsep}{5mm}{
				\begin{tabular}{c|c|c}
					\hline\hline
					$\sqrt{s}$ (GeV) &  $t$ $(\mathrm{GeV}^2)$& $\frac{d\sigma}{dt} $ $(\mu b/{\mathrm{GeV}}^2)$ \\
					\hline
					2.30&1.135&80.655$\pm$24.405\\
					\hline
					2.35&1.249&55.077$\pm$18.164\\
					\hline
					2.40&1.366&37.173$\pm$13.104 \\
					\hline
					2.45&1.485&25.063$\pm$9.296\\
					\hline
					2.50&1.607&17.010$\pm$6.650\\
					\hline
					2.55&1.732&11.673$\pm$4.640\\
					\hline
					2.60&1.859&8.121$\pm$3.306\\
					\hline
					2.65&1.989&5.740$\pm$2.381\\
					\hline
					2.70&2.121&4.127$\pm$1.738\\
					\hline
					2.75&2.256&3.021$\pm$1.288\\
					\hline
					2.80&2.393&2.250$\pm$0.970\\
					\hline\hline
			\end{tabular}}
		}
	\end{table}
	\begin{table}[h]
		\renewcommand\arraystretch{1.5} 
		\caption{The cross sections of $\pi^-p\rightarrow K^{*}\Sigma$ at $\theta_{\mathrm{c.m.}}=90^\circ$ for the case of the $P_{s\bar{s}}$ with $J^P$ = $\frac{1}{2}^-$.}
		\label{121}{\footnotesize \centering
			%\begin{tabular}{\linewidth}{ccc}
			\setlength{\tabcolsep}{5mm}{
				\begin{tabular}{c|c|c}
					\hline\hline
					$\sqrt{s}$ (GeV) &  $t$ (GeV$^2$)& $\frac{d\sigma}{dt} $ $(\mu b/{\mathrm{GeV}}^2)$  \\
					\hline
					2.30&1.135&104.376$\pm$14.620\\
					\hline
					2.35&1.249&71.407$\pm$8.284\\
					\hline
					2.40&1.366&48.478$\pm$5.898 \\
					\hline
					2.45&1.485&33.337$\pm$4.257\\
					\hline
					2.50&1.607&23.150$\pm$3.098\\
					\hline
					2.55&1.732&16.212$\pm$2.267\\
					\hline
					2.60&1.859&11.444$\pm$1.667\\
					\hline
					2.65&1.989&8.144$\pm$1.230\\
					\hline
					2.70&2.121&5.844$\pm$0.912\\
					\hline
					2.75&2.256&4.230$\pm$0.760\\
					\hline
					2.80&2.393&3.122$\pm$0.509\\
					\hline\hline
			\end{tabular}}
		}
	\end{table}
	
\begin{figure}[htbp]
		\centering
		\includegraphics[scale=0.46]{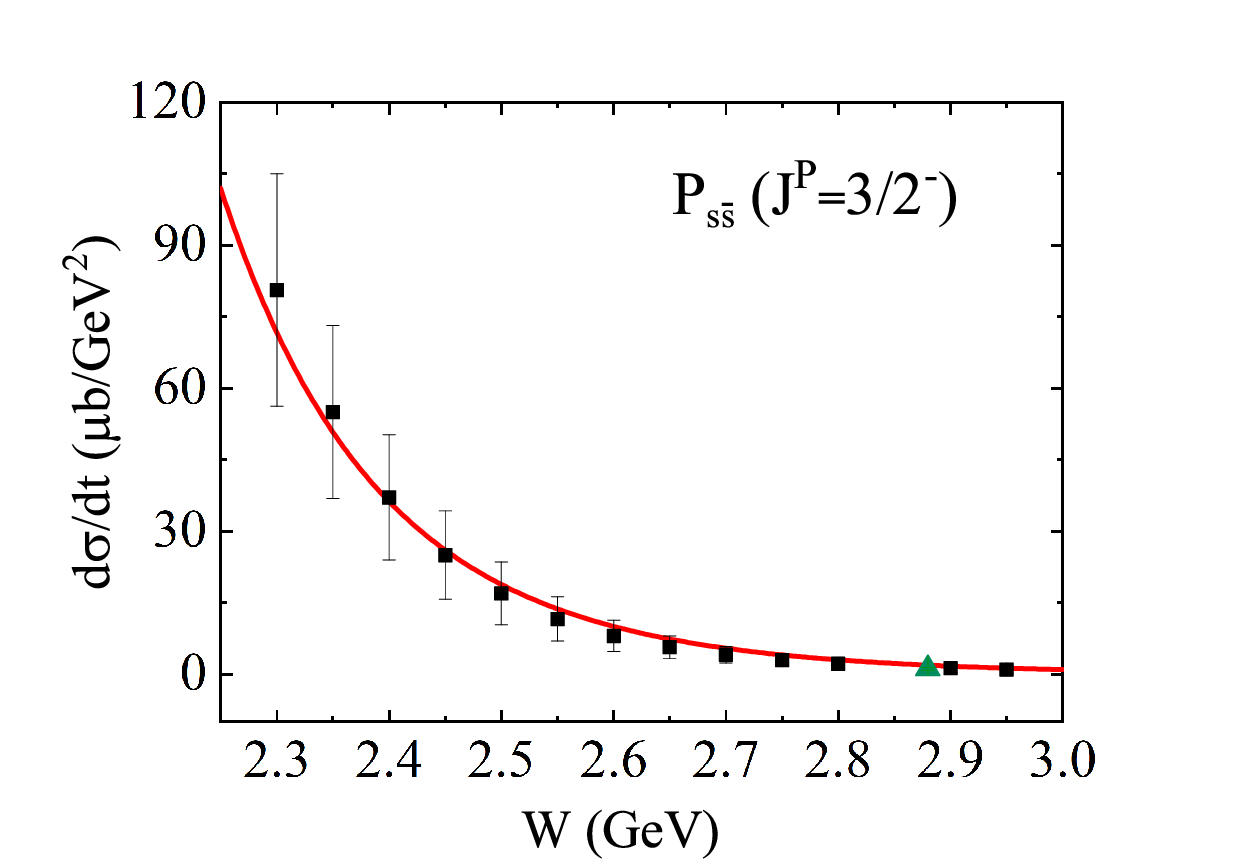}
		\caption{Differential cross section of $\pi^-p\rightarrow K^{*}\Sigma$, $d\sigma/dt$, at large meson production angle $\theta=90^\circ$ in c.m. as a function of $W$ $(W=|\sqrt{s} |)$. The green triangle and the black squares represent the experimental point and the numerical points in \cref{32}, respectively. The band stands for the error bar of the fitting parameters in \cref{zfjsgz}.}
		\label{zfjs32}
	\end{figure}

\begin{figure}[htbp]
		\centering
		\includegraphics[scale=0.46]{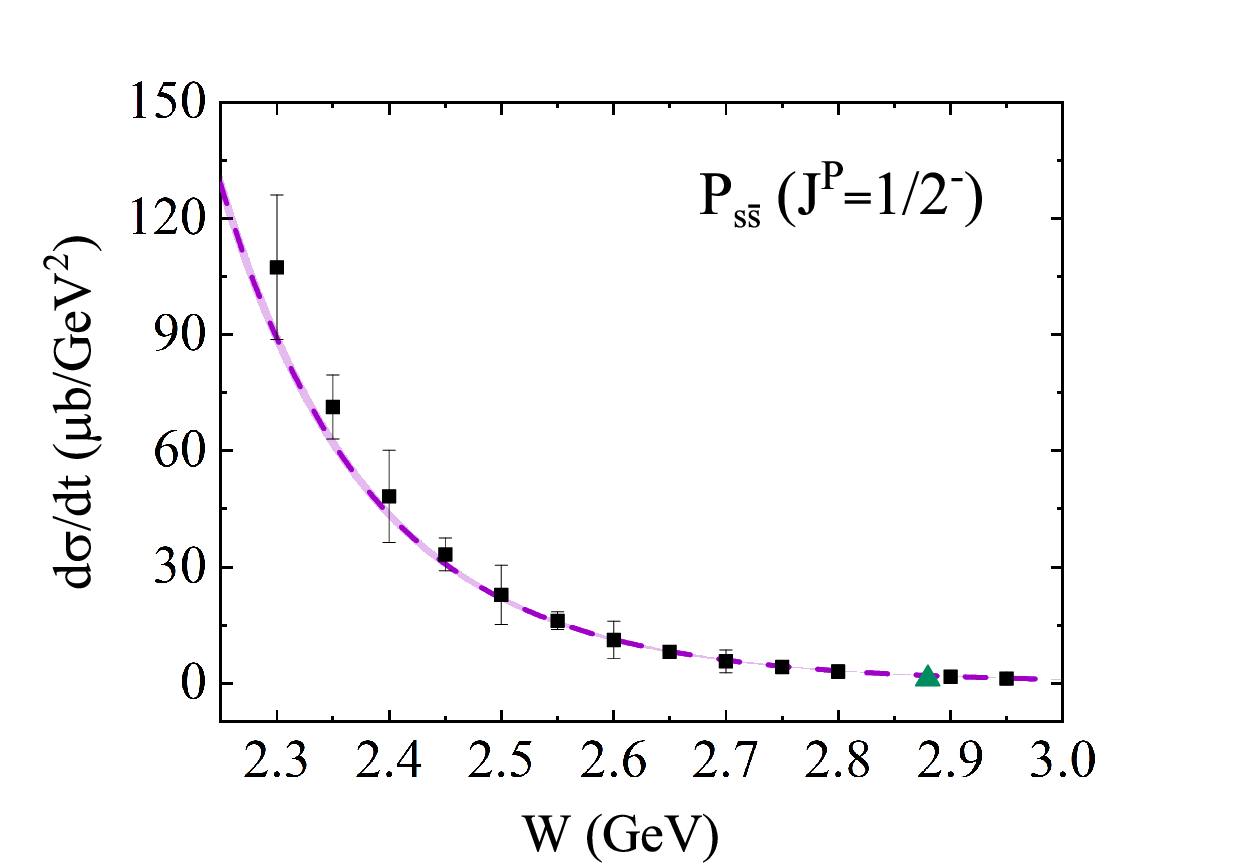}
		\caption{Differential cross section of $\pi^-p\rightarrow K^{*}\Sigma$, $d\sigma/dt$, at large meson production angle $\theta=90^\circ$ in c.m. as a function of $(W=|\sqrt{s} |)$. The green triangle and the black squares represent the experimental point and the numerical points in \cref{121}, respectively. The band stands for the error bar of the fitting parameters in \cref{zfjsgz}.}
		\label{zfjs12}
	\end{figure}

According to the rule, in the $\pi^-p \rightarrow K^*\Sigma$ reaction, the total number of constituents is $n = 2 + 3 + 2 + 3 = 10$. We calculate the cross sections for the $\pi^-p \rightarrow K^*\Sigma$ process at $\theta_{\mathrm{c.m.}}=90^\circ$, considering the cases where the $P_{s\bar{s}}$ has $J^P = \frac{3}{2}^-$ and $J^P = \frac{1}{2}^-$. The results are presented in \cref{32} and \cref{121}, respectively. By fitting the experimental data at $\sqrt{s} = 2.88$ GeV and the numerical results shown in \cref{32} and \cref{121} to the expression $d\sigma/dt = (\text{constant}) \times s^{2-n}$, the scaling factors obtained are presented in \cref{zfjsgz}. We find that the fitted values of $n$ are very close to 10, confirming the accuracy of the theoretical prediction for the cross section,. \cref{zfjs32} and \cref{zfjs12} depict the differential cross section $d\sigma/dt$ of the $\pi^-p \rightarrow K^*\Sigma$ reaction at a large meson production angle ($\theta = 90^\circ$ in the center of mass frame) as a function of $\sqrt{s}$ for the $P_{s\bar{s}}[3/2^-]$ and $P_{s\bar{s}}[1/2^-]$, respectively. Both figures clearly demonstrate that as energy increases, the theoretically predicted differential cross section aligns more closely with the constituent counting rules. This occurs because these rules are generally more applicable in high-energy, large-momentum transfer regions. In this high-energy regime, which is distant from the mass threshold of $P_{s\bar{s}}$, the contribution of $P_{s\bar{s}}$ production to the differential cross section becomes minimal. Consequently, this region does not allow the constituent counting rules to reflect the properties of the intermediate exchange particle $P_{s\bar{s}}$. Introducing the constituent counting rules serves to verify the theoretical cross-section calculations presented in this work, further supporting the reliability of our theoretical model.

	\begin{table}[h]
		\renewcommand\arraystretch{1.2} 
		\caption{Fitted values of the scaling factor $n$ by using the $\chi^2$ fitting algorithm.}
		\label{zfjsgz}{\footnotesize \centering
			%\begin{tabular}{\linewidth}{ccc}
			\setlength{\tabcolsep}{5mm}{
				\begin{tabular}{c|c|c}
					\hline\hline		 
					&  $P_{s\bar{s}}[3/2^-]$& $P_{s\bar{s}}[1/2^-]$ \\
					\hline
					$n$ &10.01$\pm$0.01&10.40$\pm$0.04\\
					\hline
					$\chi^2/d.o.f.$  &0.76&0.99\\
					\hline\hline
			\end{tabular}}
		}
	\end{table}
 
	\section{Dalitz process and experimental feasibility}\label{zhangjie5}
	\begin{figure*}[htbp]
		\centering
		\includegraphics[width=1.0\linewidth]{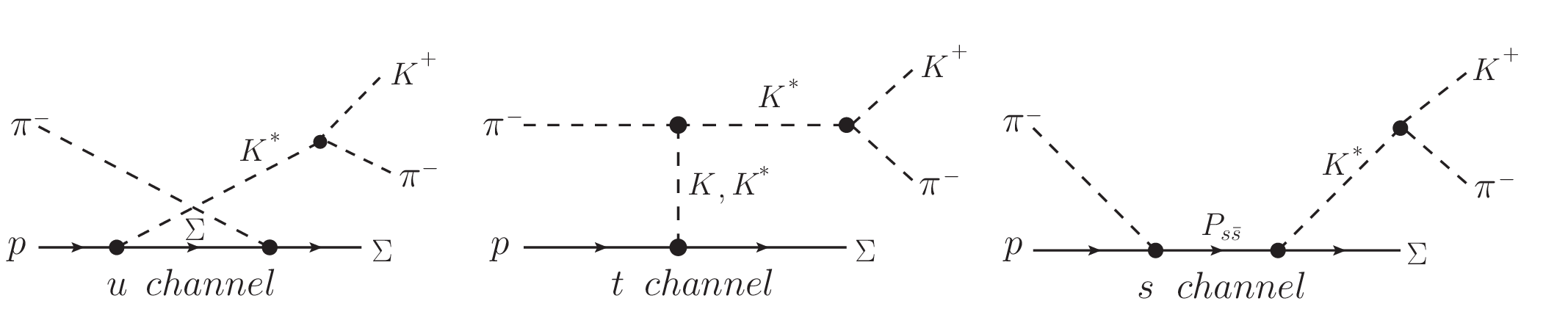}
		\caption{Feynman diagrams for the $\pi^-p\rightarrow  K^+\pi^-\Sigma^0$ reaction.}
		\label{dalitz2}
	\end{figure*}
	
\begin{figure}[htbp]
		\centering
		\includegraphics[scale=0.43]{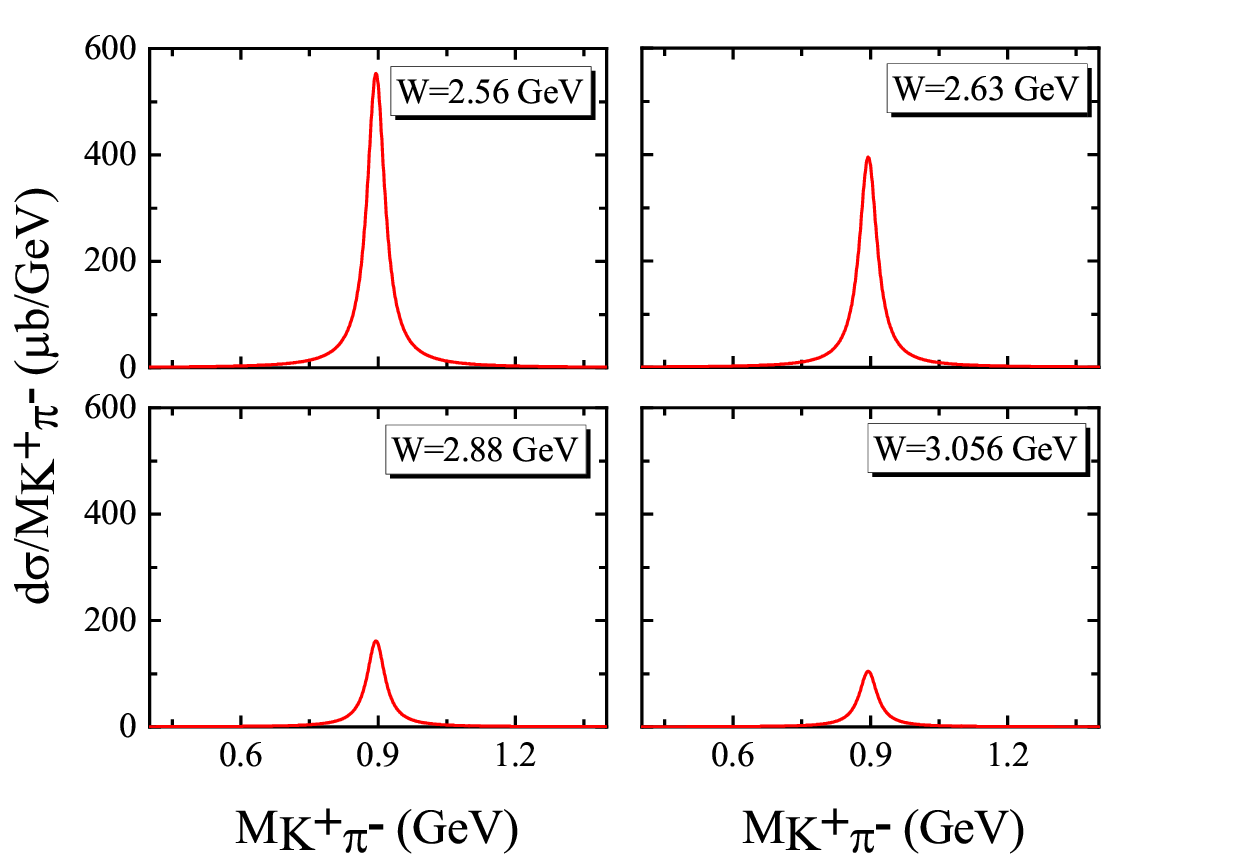}
		\caption{The invariant mass distribution $d\sigma_{\pi^-p\rightarrow K^*\Sigma\rightarrow K^+\pi^-\Sigma^0}$ /$dM_{K^+\pi^-}$ at different c.m. energies $W$ = 2.56 GeV, 2.63 GeV, 2.88 GeV, and 3.056 GeV.}
		\label{dalitz}
	\end{figure}

	In the $\pi^-p \rightarrow K^*\Sigma$ process, the $K^*$ meson produced cannot be measured directly. Instead, it must be determined by reconstructing the final particles resulting from its decay. Given that the branching ratio for $K^*$ decay to $K^+\pi^-$ is as high as $99.754\%$ \cite{ParticleDataGroup:2022pth}, analyzing the Dalitz process for $\pi^-p \rightarrow K^*\Sigma \rightarrow K^+\pi^-\Sigma^0$ is crucial for providing useful information for experimental measurements. The basic tree-level Feynman diagrams for the $\pi^-p \rightarrow K^+\pi^-\Sigma^0$ reaction are illustrated in \cref{dalitz2}. Our calculations show that the total cross section for $P_{s\bar{s}}[1/2^-]$ is slightly larger than that for the $P_{s\bar{s}}[3/2^-]$. In this study, we focus on the Dalitz process specifically for $P_{s\bar{s}}$ with $J^P = \frac{3}{2}^-$. Generally, the invariant mass spectrum of the Dalitz process is defined based on the two-body process \cite{Kim:2017nxg, Wang:2024qnk}
	\begin{eqnarray}
		\frac{d\sigma_{\pi^-p\rightarrow 
				K^* \Sigma \rightarrow K^+\pi^-\Sigma^0}}{dM_{K^+\pi^-}} \approx\frac{2M_{K^*}M_{K^+\pi^-}}{\pi}\frac{\sigma_{\pi^-p\rightarrow 
				K^* \Sigma} \Gamma_{K^*\rightarrow K^+\pi^-}}{(M^2_{K^+\pi^-}-M^2_{K^*})^2+M^2_{K^*}\Gamma^2_{K^*}},\notag
	\end{eqnarray}% 
	where $\Gamma_{K^*} = 47.3 \text{ MeV}$ and $\Gamma_{K^* \rightarrow K^+\pi^-} = 47.2 \text{ MeV}$ represent the total width and the decay width of $K^*$ to $K^+\pi^-$, respectively. The calculation results are shown in \cref{dalitz}, where peak values are observed at $M_{K^+\pi^-} = 895 \text{ MeV}$, with peaks not less than 104.57 $\mu\text{b}$/GeV. This indicates that reconstructing the $\pi^-p \rightarrow K^*\Sigma$ process through the $\pi^-p \rightarrow K^+\pi^-\Sigma^0$ process is feasible in experiments.
	
	The calculated total cross sections for $\pi^-p \rightarrow K^*\Sigma \rightarrow K^+\pi^-\Sigma^0$ and the experimental data for $\pi^-p \rightarrow K^+\pi^-\Sigma^0$ are presented in Table \ref{width1}. To assess the feasibility of experimentally detecting the $K^*$ via $\pi^-p$ scattering, we calculate the ratio $\sigma(\pi^-p \rightarrow K^*\Sigma \rightarrow K^+\pi^-\Sigma^0) / \sigma(\pi^-p \rightarrow K^+\pi^-\Sigma^0)$. The minimum ratio reaches 29.67\% at center-of-mass energies between 2.56 and 2.88 GeV, indicating that the total cross section and event number for the $\pi^-p \rightarrow K^*\Sigma$ process reconstructed via $\pi^-p \rightarrow K^+\pi^-\Sigma^0$ are sufficient for experimental measurements. Additionally, the ratio can reach up to 76.61\% at a center-of-mass energy of 2.56 GeV, with the $s$-channel $P_{s\bar{s}}$ exchange contributing 73.19\%, approximately 30.28 $\mu\text{b}$. These results suggest that a significant number of the $P_{s\bar{s}}$ can be produced in the $\pi^-p \rightarrow K^*\Sigma \rightarrow K^+\pi^-\Sigma^0$ process, providing strong support for future experimental efforts.
	\begin{table}[h]
		\renewcommand\arraystretch{1.5} 
		\caption{The total cross section of $\pi^-p \rightarrow K^+\pi^-\Sigma^0$ and $\pi^-p \rightarrow K^{*}\Sigma\rightarrow K^+\pi^-\Sigma^0$. Here, the $\sigma_1$ represents experimental cross section datas of $\pi^-p \rightarrow K^+\pi^-\Sigma^0$~\cite{Dahl:1967pg,Abramovich:1972rq}, while $\sigma_2$ is the calculated cross section of $\pi^-p \rightarrow K^{*}\Sigma\rightarrow K^+\pi^-\Sigma^0$. }
		\label{width1}{\footnotesize \centering
			%\begin{tabular}{\linewidth}{ccc}
			\setlength{\tabcolsep}{5mm}{
				\begin{tabular}{c|c|c|c}
					\hline\hline
					$W$ $(GeV)$& $\sigma_1$ $(\mu b)$ &  $\sigma_2$ $(\mu b)$ &($\sigma_2/\sigma_1$)\\
					\hline
					2.56&54&41.37&76.61\%\\
					\hline
					2.60&46&33.93&73.76\%\\
					\hline
					2.63&41&29.53&72.02\%\\
					\hline
					2.86&43&12.76&29.67\%\\
					\hline
					2.88&40&12.05&30.12\%\\
					\hline\hline
			\end{tabular}}
		}
	\end{table}
	
	As a leading hadron experimental facility, J-PARC currently provides the highest-intensity meson beams (both $K$ and $\pi$ mesons) in the several GeV energy range, making it an outstanding platform for particle and nuclear physics research \cite{Aoki:2021cqa}. Theoretical studies and experimental evidence suggest that certain light-flavored baryons may contain a significant strange quark pair ($s\bar{s}$) component, potentially forming a hidden-strange pentaquark state, such as the $P_{s\bar{s}}$. However, the existence of a hidden-strange pentaquark state has not yet been confirmed experimentally, underscoring the need for collaborative efforts between experimental and theoretical research. If J-PARC can measure the $\pi^-p \rightarrow K^*\Sigma$ process, the findings could aid in the discovery of the hidden-strange pentaquark state $P_{s\bar{s}}$ and provide valuable insights into quark interactions within hadrons, particularly the internal dynamics of diquarks.
	
	Additionally, HIAF offers high-intensity proton and ion beam currents across a wide energy range. Upon completion, HIAF will deliver the highest peak-intensity medium and low-energy heavy-ion beam currents compared to similar facilities worldwide \cite{HIAF}. If a meson beam experimental device is developed at HIAF in the future, it could enable high-precision measurements of the $\pi^-p \rightarrow K^*\Sigma$ process, further supporting the search for the $P_{s\bar{s}}$ \cite{wl:2024ext}.
	
	\section{Summary and prospect} \label{zhangjie6}
	In recent years, the $P_{s\bar{s}}$ has been considered the hidden-strange partner of the hidden-charm molecular pentaquark $P_c(4457)$, as its mass lies just below the $K^*\Sigma$ threshold \cite{He:2017aps, Ben:2023uev, Xie:2010yk, Lin:2018kcc}. In this study, we investigate the production of the hidden-strange molecular-type pentaquarks $P_{s\bar{s}}$ through the $\pi^-p$ scattering process using the effective Lagrangian approach. Specifically, we consider $K$ and $K^*$ meson exchanges in the $t$-channel, $\Sigma$ exchange in the $u$-channel, and the contribution of the $P_{s\bar{s}}$ in the $s$-channel. By fitting the total and differential cross sections of the $\pi^-p \rightarrow K^*\Sigma$ process, we demonstrate that the $P_{s\bar{s}}$ exchange provides the dominant contribution to the total cross section near the threshold energy region, offering potential insights into detecting the $P_{s\bar{s}}$ through the $\pi^-p$ scattering process. However, the lack of experimental data near the threshold imposes limitations on determining the properties of the $P_{s\bar{s}}$ via this process.
	
	Given the significant branching ratio of $K^*$ decay into $K^+\pi^-$, we calculate the Dalitz process for $\pi^-p \rightarrow K^*\Sigma \rightarrow K^+\pi^-\Sigma^0$ and compare it with experimental data for $\pi^-p \rightarrow K^+\pi^-\Sigma^0$. The results suggest that it is feasible to observe the $P_{s\bar{s}}$ through the $\pi^-p \rightarrow K^*\Sigma$ process in experiments. Notably, our calculations consider two scenarios for the $J^P$ quantum number of the $P_{s\bar{s}}$, either $\frac{1}{2}^-$ or $\frac{3}{2}^-$. 
 A major source of uncertainty in our model stems from the lack of experimental data on the partial widths of $P_{s\bar{s}}$ decays to $\pi N$. This necessitates treating the coupling constants $g_{\pi NP^*}^{3/2^-}$ and $g_{\pi NP^*}^{1/2^-}$ as free parameters, determined through fitting. In these fitting calculations, both the coupling constants and the cutoff parameters in the form factors are set as free parameters, which is an inherent aspect of this theoretical approach. Consequently, while we can calculate and analyze the production mechanism and cross section of the $P_{s\bar{s}}$ state via the $\pi^- p \rightarrow K^{*} \Sigma$ process, we cannot further deduce the internal structural properties of the $P_{s\bar{s}}$ state.
	Additionally, the application of the constituent counting rule in hard exclusive reactions is discussed. For the $\pi^-p \rightarrow K^*\Sigma$ reaction, our theoretical cross section predictions align well with the constituent counting rules, providing a strong validation of the overall cross-section calculation.
	
	Currently, high-precision measurements of the meson-nucleus scattering process can be conducted by experiments at J-PARC \cite{Aoki:2021cqa}, AMBER \cite{Adams:2018pwt}, and future facilities such as HIKE \cite{HIKE:2023ext} and HIAF \cite{HIAF}. We recommend that these experiments perform precise measurements of the $\pi^-p \rightarrow K^*\Sigma$ process, particularly the differential cross section at the forward angle ($\theta = 0^\circ$) and the $t$-distribution cross section at large momentum transfer ($\theta = 90^\circ$). This is crucial for elucidating the mechanism of the $\pi^-p \rightarrow K^*\Sigma$ process and determining the contribution of the $P_{s\bar{s}}$ state. To date, pentaquark-like states containing strange quarks have not been experimentally confirmed, necessitating further clarification through collaborative experimental and theoretical efforts. Analyzing the existence and contribution of pentaquark-like states with strange quarks via the $\pi^-p \rightarrow K^*\Sigma$ reaction is an effective approach \cite{wl:2024ext}. Moving forward, we plan to systematically investigate the contribution of excited baryon states in various $\pi^-p$ or $K^-p$ scattering processes, providing essential theoretical support for future experiments.
	
	This study is part of our ongoing efforts. The experimental beam currents and related data necessary for specific hadron spectroscopy studies can only be determined through Monte Carlo simulations based on the parameters (such as luminosity, etc.) of experimental facilities like HIAF. We will also collaborate closely with experimentalists to conduct further in-depth research in this area.
	
	\section{Acknowledgments}\label{zhangjie7}
	This work is supported by the National Natural Science Foundation of China under Grants No. 12065014, No. 12047501 and No. 12247101, the Natural Science Foundation of Gansu province under Grant No. 22JR5RA266, and the West Light Foundation of The Chinese
	Academy of Sciences under Grant No. 21JR7RA201. X.L. is also supported by National Natural Science Foundation of China under Grant No. 12335001, National Key Research and Development Program of China under Contract No. 2020YFA0406400, the 111 Project under Grant No. B20063, the Fundamental Research Funds for the Central Universities, and the project for
	top-notch innovative talents of Gansu province.

\end{document}